\documentclass[aps,amsfonts,amsmath,prd,preprint,article,nofootinbib]{revtex4}

\usepackage[toc,page]{appendix}
\newcommand{\beq}{\begin{equation}}
\newcommand{\eeq}{\end{equation}}
\newcommand{\p}{\partial}
\usepackage{amsfonts,amsmath,amssymb,epsfig}
\usepackage{subcaption}
\usepackage{hyperref} 
 \hypersetup{ 
     colorlinks=true, 
     linkcolor=black, 
     filecolor=black, 
     citecolor = blue,       
     urlcolor=blue, 
     } 
\usepackage{csquotes}
 
\numberwithin{equation}{section}

\newcommand{\nocontentsline}[3]{}
\newcommand{\tocless}[2]{\bgroup\let\addcontentsline=\nocontentsline#1{#2}\egroup}

\begin{document}
\title{Jackiw-Teitelboim and Kantowski-Sachs quantum cosmology}

\author{Georgios Fanaras and Alexander Vilenkin}

\address{ Institute of Cosmology, Department of Physics and Astronomy,\\ 
Tufts University, Medford, MA 02155, USA}

\begin{abstract}

We study quantum cosmology of the $2D$ Jackiw-Teitelboim (JT) gravity with $\Lambda>0$ and calculate the Hartle-Hawking (HH) wave function for this model in the minisuperspace framework.  Our approach is guided by the observation that the JT dynamics can be mapped exactly onto that of the Kantowski-Sachs (KS) model describing a homogeneous universe with spatial sections of $S^1\times S^2$ topology.  This allows us to establish a JT-KS correspondence between the wave functions of the models. 
We obtain the semiclassical Hartle-Hawking wave function by evaluating the path integral with appropriate boundary conditions and employing the methods of Picard-Lefschetz theory. The JT-KS connection formulas allow us to translate this result to JT gravity, define the HH wave function and obtain a probability distribution for the dilaton field.

\end{abstract}  

\maketitle

  \tableofcontents

\section{Introduction}

In quantum cosmology the entire universe is treated quantum mechanically and is described by a wave function, rather than by a classical spacetime. The wave function $\Psi(g,\phi)$ is defined on the space of all 3-geometries ($g$) and matter field configurations ($\phi$), called superspace.  It can be found by solving the Wheeler-DeWitt (WDW) equation
\beq
{\cal H}\Psi=0,
\eeq
where ${\cal H}$ is the Hamiltonian operator.  Alternatively, one can consider the transition amplitude from the initial state $(g',\phi')$ to the final state $(g,\phi)$, which can be expressed as a path integral,
\beq
G(g,\phi|g',\phi') = \int_{(g',\phi')}^{(g,\phi)} ~ e^{\textstyle{iS}},
\label{pathint}
\eeq
where $S$ is the action  and the integration is over the histories interpolating between the initial and final configurations.  In general, $G$ is a Green's function of the WDW equation \cite{Teitelboim}.  
But if $(g',\phi')$ is at the boundary of superspace, or if the geometries that are being integrated over have a single boundary at $g$, then $G$ is a solution of the WDW equation 
and the path integral (\ref{pathint}) may be used to define a wave function of the universe.

The choice of the boundary conditions for the WDW equation and of the class of paths included in the path integral representation of $\Psi$ has been a subject of ongoing debate. The most developed proposals in this regard are the Hartle-Hawking \cite{HH} and the tunneling \cite{Vilenkin:1987kf,Vilenkin:1994rn} wave functions.\footnote{For early work closely related to HH and tunneling proposals, see Refs.~\cite{Vilenkin:1982de} and \cite{Linde:1983mx,Rubakov:1984bh,Vilenkin:1984wp,Zeldovich:1984vk} respectively.}
The intuition behind both of these proposals is that the universe originates `out of nothing' in a non-singular way.  But despite a large amount of work, a consensus on the precise definition of these wave functions has not yet been reached.  {In fact, the two wave functions are often confused with one another.} 

 The Hartle-Hawking (HH) wave function is usually defined in terms of a Euclidean path integral,
\beq
\Psi_{HH}(g,\phi) = \int ^{(g,\phi)}  e^{\textstyle{-S_E}},
\label{HH0}
\eeq
where $S_E$ is the Euclidean action and the integration is over regular compact geometries with a single boundary on which the boundary values $(g,\phi)$ are specified.  The tunneling wave function is defined either by an outgoing-wave boundary condition in superspace or by a path integral over Lorentzian histories interpolating between a vanishing 3-geometry and the configuration $(g,\phi)$.  Here we will focus on the HH wave function; the tunneling wave function will be discussed in a separate publication.

In the last few years there has been a renewed interest in quantum cosmology, inspired by recent work on the exactly soluble $(1+1)$-dimensional quantum gravity model -- the Jackiw-Teitelboim (JT) gravity \cite{Jackiw:1984je,Teitelboim:1983ux}.  This model can be thought of as a quantum theory of a one-dimensional closed universe.  Apart from the scale factor $a$, it also includes an evolving scalar field $\phi$ -- the dilaton, which makes a comparison with higher-dimensional models somewhat less informative.  On the positive side, one can hope that exact solubility of the model may provide new insights into the nature of the wave function of the universe.\footnote{ An exact quantization of JT model was first developed by Henneaux \cite{Henneaux}.  For recent discussions see Refs.~\cite{Maldacena,Iliesiu,Trivedi,Stanford} and references therein.} 

 The HH wave function for JT gravity has been recently discussed in the interesting paper by 
Maldacena, Turiaci and Yang (MTY) \cite{Maldacena}.   They calculated the wave function in the leading semiclassical order in the limit of large a and included the pre-exponential factor suggested by the Schwarzian analysis. In difference from the Hartle and Hawking approach, MTY focused on the outgoing branch of the wave function, describing expanding universes at large a.  As a result the asymptotic behavior of $\Psi$ is more consistent with the tunneling boundary conditions.

Another interesting recent work is the paper by Iliesiu, Kruthoff, Turiaci and Verlinde (IKTV) \cite{Iliesiu}.
They presented an exact solution to the WDW equation of JT gravity, which they interpreted as the HH wave function, but their choice of boundary conditions was different from the earlier literature.  Hartle \& Hawking and most of the subsequent authors required that geometries included in the path integral close off smoothly in the limit of small universes.  Instead, IKTV imposed a boundary condition in the opposite limit, requiring that the wave function exhibits Schwarzian behavior when the universe is large.  The assumption of regularity and closure is implicit in their discussion, but these conditions are not explicitly enforced.  The resulting wave function agrees with the semiclassical analysis of MTY in the appropriate limit.  However, IKTV note an unexpected feature: the wave function develops a strong singularity at a finite value of the scale factor.

 In the present paper we take a different approach to JT quantum cosmology.  It is based on the observation of MTY that JT model can be obtained from $4D$ gravity by dimensional reduction.  We shall use this connection between $2D$ and $4D$ theories as a guide to defining the cosmological HH wave function in the JT model.  In their paper MTY discussed a dimensional reduction from a nearly extremal Schwarzschild-de Sitter solution, with the extra two dimensions compactified on a sphere (see also \cite{Fabbri,Bousso} for earlier work).    Since our emphasis is on the cosmological aspects of the theory, we find it more useful to consider a cosmological $4D$ model describing a homogeneous universe with spatial sections having $S^1\times S^2$ topology, known as the Kantowski-Sachs model.  The main difference from the MTY and IKTV work is that we impose the boundary conditions in the small universe limit, requiring that the geometry closes off in a regular way.

 We begin in the next section by reviewing JT gravity and its quantization, discussing in particular the semiclassical analysis of MTY and the exact solutions of IKTV.  We argue that these solutions are not suitable to represent the HH wave function.  We also discuss how JT model can be obtained by dimensional reduction from $4D$ gravity.

 In Section 3 we review the quantum cosmology of the Kantowski-Sachs (KS) model, following largely the treatment of Halliwell and Louko (HL) \cite{HL}.  We establish an exact correspondence between the WDW equations for KS and JT models.  Furthermore, we show that the transition amplitude between states with specified initial and final scale factors calculated by HL is closely related to the wave function found by IKTV.  It follows from this analysis that their wave function satisfies an equation with a singular source and thus is not a solution of the WDW equation.  This accounts for the divergence of the wave function pointed out by IKTV.

The semiclassical HH wave function for the KS model is discussed in Section 4.  HL studied this wave function only for a vanishing cosmological constant, $\Lambda=0$.  Here we will need to extend their analysis to the case of $\Lambda>0$, which is significantly more complicated.  We impose the boundary conditions of smooth closure in the limit of small universes and follow standard methods to reduce the problem to evaluation of a lapse ($N$) integral over some contour ${\cal C}$ in the complex $N$ plane.  The choice of the contour ${\cal C}$ is restricted by the requirements that the HH wave function is expected to satisfy.  We argue that there is only a single acceptable choice, with all other acceptable choices equivalent to it. 

In the semiclassical limit the dominant contribution to the integral is given by saddle points of the action.  We find these saddle points, as well as the steepest descent and ascent lines, and use the Picard-Lefschetz prescription to deform the contour so that the integral becomes absolutely convergent.  The integral is then evaluated in the WKB approximation
for the range of parameters most relevant for the connection to JT.

 In Section 5 we use the HH wave function calculated in the preceding section to find the probability distribution for the radius of $S^2$ at a given radius of $S^1$ in our $S^1\times S^2$ model.  In Section 6 we use the correspondence between JT and KS models to define the HH wave function in JT gravity.  We use this wave function to determine the probability distribution for the dilaton field $\phi$.
Our results are summarized and discussed in Section 7.  Some technical details are relegated to the Appendix.

\section{JT gravity}

\subsection{The action}

The action of the JT model is \cite{Jackiw:1984je,Teitelboim:1983ux}
\beq
S=\int d^2 x\sqrt{-g} \phi (R-2H^2) -2\int_B \phi_b K,
\label{JTaction}
\eeq
where $R$ is the $2D$ spacetime curvature, $H={\rm const}$, $\phi$ is the dilaton field, $\phi_b$ is its value at the boundary, and $K$ is the extrinsic curvature of the boundary curve $B$. Throughout the paper we shall assume that $H>0$. Variation with respect to $\phi$ yields $R=2H^2$, telling us that the $2D$ spacetime is a de Sitter space with expansion rate $H$.

With the metric represented as
\beq
ds^2=-N^2 dt^2+a^2 dx^2, 
\eeq
where $0<x<2\pi$ and $N$ is the lapse function, the state vector is a functional 
\beq
\Psi[a(x),\phi(x)].  
\label{functional}
\eeq
We can choose the gauge so that $\phi_b={\rm const}$ at the boundary.  Furthermore, we are going to adopt a minisuperspace picture, where $a=a(t)$, independent of $x$, and the boundary is a circle, $t={\rm const}$.  Then $\Psi$ is an ordinary function $\Psi(a,\phi)$.  It has been shown in \cite{Iliesiu} that due to the simplicity of the model, the wave functional (\ref{functional}) can be recovered from the minisuperspace wave function $\Psi(a,\phi)$.  Here, we shall restrict our analysis to the minisuperspace model with $a=a(t),~\phi=\phi(t)$.

After integration by parts the action (\ref{JTaction}) can be represented as 
\beq
S=-4\pi\int dt\left(\frac{{\dot a}{\dot\phi}}{N}+NH^2 a\phi\right),
\eeq
where dots stand for derivatives with respect to $t$.
We are going to use the gauge $N={\rm const}$.  The momenta conjugate to $a$ and $\phi$ are
\beq
\Pi_a=-\frac{4\pi}{N}{\dot\phi},~~~~ \Pi_\phi=-\frac{4\pi}{N}{\dot a}.
\label{Pis}
\eeq

The equations of motion obtained by varying the action with respect to $a$ and $\phi$ are
\beq
{\ddot a}-H^2 a=0,
\eeq
\beq 
{\ddot\phi}-H^2\phi=0,
\eeq
where we have set $N=1$.
The Hamiltonian constraint is obtained by varying with respect to $N$:
\beq
{\dot a}{\dot\phi}=H^2 a\phi,
\label{WDWconstraint1}
\eeq
or
\beq
\Pi_a\Pi_\phi=16\pi^2 H^2 a\phi.
\label{WDWconstraint}
\eeq
The classical solution of these equations is 
\beq
a=a_0 \cosh (Ht), ~~~~ \phi=\phi_0 \sinh (Ht)
\label{solution}
\eeq
with $a_0, \phi_0={\rm const}$.  We shall set $a_0=H^{-1}$, so that the metric covers the full de Sitter space in a nonsingular way.

\subsection{Semiclassical wave function}

To lowest order in the WKB approximation, the wave function is given by
\beq
\Psi\sim e^{iS_{cl}},
\label{PsiS}
\eeq
where $S_{cl}$ is the classical action,
\beq
S_{cl}=  -2\int_0^{2\pi} dx~ a\phi_b K=-4\pi a\phi_b K.
\label{SK}
\eeq
Here, we used the fact that $R=2H^2$ in the classical solution, so only the surface terms make a contribution, and that $\phi$ and $K$ are constant on the boundary.
Following the no-boundary philosophy, we assume that the (Euclideanized) geometry closes off smoothly, so that there is no boundary contribution at $a=0$.

In the classically allowed region $(a>H^{-1})$, the extrinsic curvature $K$ is given by
\beq
K=\frac{\dot a}{a} =H\tanh (Ht) = a^{-1}\sqrt{H^2 a^2 -1},
\eeq
where we have used Eq.~(\ref{solution}) with $a_0=H^{-1}$.  Substituting this in Eqs.~(\ref{SK}) and (\ref{PsiS}), we obtain
\beq
\Psi\propto \exp\left(-4\pi i\phi_b\sqrt{H^2 a^2-1}\right)
\label{Psi}
\eeq
A linearly independent WKB wave function is a complex conjugate of (\ref{Psi}).  A general WKB solution is a linear combination of the two.  The semiclassical approximation applies when the action is large, $\phi_b\sqrt{H^2 a^2-1}\gg 1$.

The momentum operator $\Pi_\phi$ acting on $\Psi$ gives
\beq 
\Pi_\phi\Psi =-i\p_\phi\Psi =-4\pi\sqrt{H^2 a^2-1}\Psi.
\eeq
The classical momentum is given by Eq.~(\ref{Pis}), so we get 
\beq
{\dot a}=\sqrt{H^2 a^2-1}.
\eeq
This agrees with the expanding branch of the classical solution (\ref{solution}).  The complex conjugate wave function describes a contracting universe.

\subsection{Exact solutions of the WDW equation}

The WDW equation corresponding to the Hamiltonian constraint (\ref{WDWconstraint1}) is
\beq
(\p_a\p_\phi+16\pi^2 H^2 a\phi){\tilde\Psi} =0.
\label{WDW5}
\eeq
Here, 
\beq
{\tilde\Psi}=\Psi/a
\eeq 
and the factor $1/a$ comes from the factor ordering indicated by the exact quantization of the JT model by Henneaux \cite{Henneaux} (see IKTV \cite{Iliesiu} for a detailed explanation).
Following MTY \cite{Maldacena}, we introduce new variables
\beq
u=\phi^2, ~~~ v=(2\pi)^2 (H^2 a^2-1).
\eeq
Then the WDW equation becomes
\beq
(\p_u\p_v+1){\tilde\Psi}=0.
\label{uvWDW}
\eeq
Yet another change of variables
\beq
T=\sqrt{uv},~~~ \xi=\frac{1}{2}\ln\frac{v}{u}
\eeq
brings the equation to a separable form
\beq
-\frac{1}{T}\p_T(T\p_T{\tilde\Psi})+\frac{1}{T^2}\p_\xi^2{\tilde\Psi}-4{\tilde\Psi}=0.
\eeq

With the ansatz
\beq
{\tilde\Psi}_m=e^{m\xi}f_m(T)
\label{WDWsoln}
\eeq
we obtain an equation for $f_m(T)$:
\beq
{f_m}'' +\frac{1}{T}{f_m}'-\frac{m^2}{T^2}f_m+4f_m=0.
\eeq
The solution is
\beq
f_m(T)=Z_m(2T),
\label{Zm}
\eeq
where $Z_m$ is a Bessel function.  

Following MTY, IKTV required that the wave function should describe an expanding universe in the limit of large $a$.  Then the appropriate choice of Bessel functions is $H_m^{(2)}(2T)$.  The general solution of the WDW equation is then a linear combination of functions of the form
(\ref{WDWsoln}):
\beq
{\tilde\Psi}_m=\left(\frac{v}{u}\right)^{m/2} H_m^{(2)} (2T).
\eeq
In terms of the variables $a$ and $\phi$, the argument of the Bessel functions is
\beq
2T=4\pi\phi\sqrt{H^2 a^2-1}.
\eeq
The asymptotic form of the Bessel functions at large $T$ is $H_m^{(2)}(2T)\propto T^{-1/2}e^{-2iT}$; hence 
\beq
{\Psi}_m(Ha\phi\gg 1)\propto \left(\frac{a}{\phi}\right)^{m+1/2}\exp\left(-4\pi i \phi\sqrt{H^2 a^2-1}\right),
\label{Psim2}
\eeq
where we 
have accounted for the factor $1/a$ relating $\Psi_m$ and ${\tilde\Psi}_m$.
All these functions have the same asymptotic exponential factor as the WKB wave function (\ref{Psi}).
So in order to choose between them one has to determine the pre-exponential factor.

In the path integral formulation, the semiclassical pre-factor is determined by quantum fluctuations about the classical solution.  In the JT model these are fluctuations in the shape of the boundary curve, which are described by the Schwarzian theory and yield a one-loop pre-factor $(\phi/a)^{3/2}$ at large $a$ \cite{StanfordWitten}.  It is shown in \cite{StanfordWitten} that this result is one-loop exact, so there are no further corrections.  This pre-factor is obtained by setting $m=-2$ in (\ref{Psim2}).  Then the exact wave function takes the form
\beq
\Psi(a,\phi)=\frac{a\phi^2}{H^2 a^2-1}H_2^{(2)}\left(4\pi\phi\sqrt{H^2 a^2-1}\right) ~~~~ (Ha>1).
\label{PsiMI}
\eeq
Analytic continuation of this wave function to $Ha<1$ is not unique because of the singularity at $Ha=1$.  IKTV specify the wave function in the entire range of $a$ by replacing the Hankel function in (\ref{PsiMI}) with $K_2\left(i\sqrt{\phi^2(H^2 a^2 -1-i\epsilon)}\right)$, which gives\footnote{{The inclusion of the term $i\epsilon$ with $\epsilon\to +0$ is needed to make the solution well-defined on the branch cut.}}
\beq
\Psi(a,\phi)=\frac{2i}{\pi}\frac{a\phi^2}{H^2 a^2-1}K_2 \left(4\pi\phi\sqrt{H^2 a^2-1}\right) ~~~~ (Ha<1).
\label{PsiMI2}
\eeq

IKTV identify the wave function (\ref{PsiMI}),(\ref{PsiMI2}) with the HH wave function for JT gravity.  There are however some problems with this identification.  We first note that one of the defining properties of the HH wave function is that it is real.  This can be interpreted as reflecting the CPT invariance of the HH state \cite{HHH}.  On the other hand, the tunneling wave function is specified by the outgoing wave boundary condition, which in the present context requires that the large $a$ asymptotic of $\Psi$ should only include terms corresponding to expanding universes.  This seems to suggest that the wave function described by Eqs.(\ref{PsiMI}),(\ref{PsiMI2}) is more appropriately interpreted as the tunneling wave function.

More importantly, the wave function (\ref{PsiMI}) has a singularity at $a=H^{-1}$.  It is actually not a solution of the WDW equation.  We will show in Sec.III.C that it satisfies
\beq
{\cal H}\Psi\propto \delta(\phi)\delta''(a-H^{-1}).
\eeq
Hence it is not suitable for the role of HH or tunneling wave function.

IKTV have also proposed another candidate for the HH wave function:
\beq
\Psi(a,\phi)=\frac{a\phi^2}{H^2 a^2-1}J_2 \left(4\pi\phi\sqrt{H^2 a^2-1}\right).
\label{PsiI}
\eeq
This wave function is real and non-singular.  However, its behavior in the classically forbidden range $a<H^{-1}$ is very different from what is expected for the semiclassical HH wave function.  We have
\beq
\Psi(a<H^{-1})=\frac{a\phi^2}{1-H^2 a^2}I_2 \left(4\pi\sqrt{\phi^2(1-H^2 a^2)}\right)\sim\frac{a\phi^{3/2}}{\sqrt{4\pi}(1-H^2a^2)^{5/4}}\exp\left(4\pi |\phi| \sqrt{1-H^2 a^2}\right),
\eeq
where the last expression is the asymptotic form of $\Psi$ assuming that the argument of $I_2$ is large.
As $a$ varies from $a=0$ to $a\sim H^{-1}$, the exponential factor in $\Psi$ decreases, which is opposite to the expected behavior of the HH wave function.

\subsection{Dimensional reduction}

MTY discussed the relation between JT and Einstein $4D$ gravity using dimensional reduction from a nearly extremal Schwarzschild-de Sitter solution.  Since our emphasis is on the cosmological aspects of the theory, we find it more useful to consider a cosmological $4D$ model describing a universe with spatial sections having $S^1\times S^2$ topology and the metric
\beq
ds^2=-dt^2+a^2(x,t) dx^2+b^2(x,t) d\Omega^2.
\label{inhomogeneous}
\eeq
Here, $0<x<2\pi$ and $d\Omega^2$ is the metric on a unit sphere.  Substituting this in the Einstein-Hilbert action (in Planck units)
\beq
S=\frac{1}{16\pi}\int d^4 x\sqrt{-g^{(4)}}\left(R^{(4)}-2\Lambda \right),
\label{EHaction}
\eeq
where $\Lambda$ is the $4D$ cosmological constant, and integrating over the angular variables we obtain
\beq
S=\int d^2 x \sqrt{-g}\left[\frac{b^2}{4}R +\frac{1}{2}(\nabla b)^2+\frac{1}{2}-\frac{1}{2}\Lambda b^2\right].
\eeq
Here, $R$ and $g$ are respectively the $2D$ curvature scalar and the metric determinant and we have omitted the boundary term.

Following \cite{Louis-Martinez}, we can remove the gradient term in the action by a conformal rescaling
\beq
{\bar g}_{\mu\nu}=\Omega^2(b)g_{\mu\nu}
\eeq
with
\beq
\frac{d \ln\Omega}{d\ln b}=\frac{1}{2}.
\eeq
This has the solution
\beq
\Omega=(b/2)^{1/2},
\eeq
where we have chosen the normalization factor for future convenience.  The action then reduces to
\beq
S=\int d^2 x \sqrt{-{\bar g}}\left[{\bar\phi}{\bar R}-V({\bar\phi})\right],
\eeq
where ${\bar\phi}=b^2/4$ and
\beq
V({\bar\phi})=2\Lambda\sqrt{\bar\phi}-\frac{1}{2\sqrt{\bar\phi}}.
\label{V}
\eeq

We define
\beq
{\bar\phi}=\phi_0+{\phi},
\eeq
where $\phi_0=1/4\Lambda$, so that $V(\phi_0)=0$.  We shall assume that $\Lambda\ll 1$, so $\phi_0\gg 1$.
Then, for $|\phi|\ll\phi_0$ we can expand the potential (\ref{V}) around $\phi=0$.  Neglecting quadratic and higher order terms in the expansion, we obtain an approximate JT action
\beq
S\approx \phi_0\int d^2 x \sqrt{-{\bar g}} {\bar R}+\int d^2 x \sqrt{-{\bar g}}{\phi}\left({\bar R}-2{\bar\Lambda}\right),
\label{JTapprox}
\eeq
where ${\bar\Lambda}=2\Lambda^{3/2}$.

Since the second term in (\ref{JTapprox}) already includes a factor of $\phi$, we can use 
\beq
{\bar g}_{\mu\nu}\approx \frac{1}{2\sqrt{\Lambda}}g_{\mu\nu},~~~{\bar R}\approx 2\sqrt{\Lambda}R.
\label{Nariai}
\eeq
Hence, in the same approximation we can rewrite the action in terms of the original metric $g_{\mu\nu}$ and the cosmological constant $\Lambda$ as
\beq
S\approx \phi_0\int d^2 x \sqrt{-{g}} {R}+\int d^2 x \sqrt{-{g}}{\phi}\left({R}-2{\Lambda}\right),
\label{2Daction}
\eeq


The above analysis suggests that in the appropriate limit the dynamics of the $4D$ cosmological model (\ref{inhomogeneous})
is well approximated by that of the JT gravity (\ref{JTaction}) with $\Lambda=H^2$.  The radius of the sphere $S^2$ in this limit is $b\approx H^{-1}$.  We will focus on this regime in most of the paper, but in the next section we will see that in the minisuperspace setting the two models are even more closely related and can be mapped onto one another for arbitrary values of the scale factors $a$ and $b$.

\section{Kantowski-Sachs model}

\subsection{Classical dynamics}

As already mentioned, our focus in this paper will be on homogeneous minisuperspace models.  Hence we will use a homogeneous version of the model (\ref{inhomogeneous}), with the scale factors $a$ and $b$ independent of $x$, for dimensional reduction.  This is the Kantowski-Sachs (KS) model \cite{KS} describing a homogeneous universe with spatial sections of $S^1\times S^2$ topology.

Following Halliwell and Louko \cite{HL}, we represent the metric of the KS model as
\beq
ds^2=-\frac{N^2}{a^2}d\tau^2+a^2 dx^2+b^2 d\Omega^2,
\label{KSEmetric}
\eeq
where $N$, $a$ and $b$ are functions of time $\tau$, which we can choose to vary in the range $0<\tau<1$.  After substituting this in the Lorentzian Einstein-Hilbert action with a cosmological constant $\Lambda\equiv H^2$ and integrating over $x$ and over the angular variables, the action reduces to
\beq
S=-\pi\int_0^1 d\tau \left[\frac{{\dot b}{\dot c}}{N}+N(H^2 b^2-1)\right],
\eeq
where we have introduced a new variable $c=a^2 b$.

The factor $1/a^2$ is added in the first term of (\ref{KSEmetric}) in order to simplify the equations of motion, which take the form
\beq
{\ddot b}=0,
\label{1}
\eeq
\beq
\frac{\ddot c}{N^2}-2H^2 b=0,
\label{2}
\eeq
where overdots stand for derivatives with respect to $\tau$.
The constraint equation is obtained by varying the action with respect to $N$:
\beq
\frac{{\dot b}{\dot c}}{N}-N(H^2 b^2-1)=0.
\label{3}
\eeq

An important solution of these equations is obtained by setting ${\dot b}=0$.  Then Eq.(\ref{3}) tells us that $b=H^{-1}$ and Eq.(\ref{2}), expressed in terms of the proper time variable $t=\int d\tau/a(\tau)$, becomes
\beq
\frac{d^2 a}{dt^2}=H^2 a,
\eeq
which has a solution 
\beq
a(t)=H^{-1}\cosh(Ht),
\eeq
where we have set $N=1$.
This is the Nariai solution, which is a product of a $2D$ de Sitter space and a 2-sphere of radius $H^{-1}$ \cite{Nariai}.  

It follows from Eq.(\ref{1}) that ${\dot b}$ cannot change sign, indicating that the Nariai solution is unstable.  If we perturb it by giving the radius of the sphere $b$ a slight velocity, the sphere will collapse if ${\dot b}<0$ and will expand to infinite size if ${\dot b}>0$.

The Euclidean continuation of the Nariai solution is a product of two spheres of radius $H^{-1}$.  It is often referred to as the Nariai instanton and describes nucleation of extremal black holes in de Sitter space \cite{Ginsparg:1982rs,Bousso:1996au}.

\subsection{WDW equation}

The quantum cosmology of the KS model has been studied by a number of authors \cite{Laflamme,HL,Conradi,Anninos:2012ft,Hertog and Conti}.  Some exact solutions of the WDW equation have been found and semiclassical methods have been used to study more general solutions.
Here we will follow the method of Halliwell and Louko (HL) \cite{HL} which allows one not only to find the  saddle points of the action, but also helps to determine which saddle points contribute to the semiclassical wave function.  This method will also be useful for interpreting the solution (\ref{PsiMI}) found by IKTV.

The momenta conjugate to the variables $b$ and $c$ are
\beq
p_b=-\pi{\dot c}/N,~~~~ p_c=-\pi{\dot b}/N.
\eeq
Using this in the constraint equation (\ref{3}) and replacing $p_b\to-i\p/\p b$, $p_c\to -i\p/\p c$, we obtain the WDW equation 
\beq
\pi{\cal H}\Psi=\left[\p_b\p_c+\pi^2(H^2 b^2-1)\right]\Psi=0.
\label{KSWDW}
\eeq

This equation can be simplified by introducing a new variable $\xi$, which is related to $b$ as $d\xi=\pi^2(H^2 b^2-1)db$.  Choosing the integration constant so that $\xi(Hb=1)=0$, we have
\beq
\xi=\frac{\pi^2}{3H}(H^3 b^3-3Hb+2)=\frac{\pi^2}{3H}(Hb-1)^2(Hb+2).
\eeq
We also introduce the variable $\rho=c-H^{-3}$; then the WDW equation becomes
\beq
(\p_\xi \p_\rho+1)\Psi=0.
\label{xirhoWDW}
\eeq

We note that this equation has the same form as Eq.(\ref{uvWDW}) for the JT model.  The difference is that Eq.(\ref{uvWDW}) is for the function ${\tilde\Psi}=\Psi/a$, where the factor $1/a$ appeared due to a particular choice of the factor ordering.  Following the same steps as in Sec.2.3, we find that Eq.(\ref{xirhoWDW}) has exact solutions
\beq
\Psi_m=\left(\frac{\rho}{\xi}\right)^{m/2} H_m^{(2)} (2\sqrt{\xi\rho}).
\label{Psimxirho}
\eeq
The solution with $m=-2$ corresponds to the IKTV solution (\ref{PsiMI}).  We expect this solution to agree with (\ref{PsiMI}) when $b\approx H^{-1}$.  The argument of the Hankel function in (\ref{Psimxirho}) is
\beq
2\sqrt{\xi\rho}=\frac{2\pi}{H^2}\sqrt{\frac{Hb+2}{3}}(Hb-1)\sqrt{H^3 a^2 b-1}\approx\frac{2\pi}{H^2}(Hb-1)\sqrt{H^2 a^2-1}.
\eeq
Comparing this with the argument of the Hankel function in (\ref{PsiMI}), we can identify
\beq
\frac{Hb-1}{H^2}\approx 2\phi.
\eeq

It is interesting to note that the correspondence between the two wave function extends beyond this approximation.  If we define
\beq
{\tilde a}=(Hb)^{1/2}a,~~~~ {\phi} =\frac{(Hb-1)}{2H^2}\sqrt{\frac{Hb+2}{3}},
\label{tildeaphi}
\eeq
then ${\tilde a}\Psi({\tilde a},{\phi})$ exactly reproduces the wave function (\ref{PsiMI}).
 More generally, the transformation (\ref{tildeaphi}) can be used to obtain a solution to the WDW equation of the JT model from that of the KS model and {\it vice versa}.
Note also that ${\tilde a}$ and ${\phi}$ are simply related to the variables $\xi$ and $\rho$:
\beq
\xi=4\pi^{2}H^{3}{\phi}^{2},~~~~\rho=H^{-3}(H^2{\tilde a}^2-1).
\eeq

We thus see that JT and KS minisuperspace models are formally equivalent to one another.  This equivalence, however, does not extend beyond minisuperspace.  In the JT case, the minisuperspace wave function can be extended to a wave function in full superspace, but in the KS model the number of variables in the wave function and the pre-exponential factor depend on which perturbation modes are included in the minisuperspace truncation.  Equivalence between the two models at the minisuperspace level will nevertheless be sufficient for our purposes here.

\subsection{Transition amplitude}

We now consider the transition amplitude from the initial state $\{b',c'\}$ to the final state $\{b,c\}$.  We will be particularly interested in the initial state 
\beq
\{b',c'\}=\{H^{-1},H^{-3}\} 
\label{Nic}
\eeq
corresponding to the bounce point of the Nariai solution.  We shall refer to it as `Nariai initial conditions'.

The general framework for calculating transition amplitudes has been discussed by HL \cite{HL}.
For a general minisuperspace model described by the Hamiltonian
\beq
{\cal H}=\frac{1}{2}f^{\alpha\beta}(q)p_\alpha p_\beta+V(q),
\eeq
where $q^\alpha$ are the generalized coordinates and $p_\alpha$ their conjugate momenta,  
the transition amplitude between $q'$ and $q$ is
\beq
G(q|q')=\int_0^\infty dN\int \mathcal{D}p \mathcal{D}q e^{iS} = \int_0^\infty dN \langle q,N|q',0\rangle.
\label{G}
\eeq
Here $N$ is the lapse parameter, the action is
\beq
S=\int_0^1 d\tau \left(p_\alpha {\dot q}^\alpha-N{\cal H}\right)
\eeq
 and the path integral is over histories interpolating between $q'$ and $q$.

HL show that the amplitude (\ref{G}) satisfies
\beq
{\cal H}G(q|q')=-i\langle q,0|q',0\rangle=-i\delta(q,q'),
\label{HG}
\eeq
where 
\beq
{\cal H}=-\frac{1}{2}\nabla^2+\zeta R+ V
\eeq
is the Hamiltonian operator, $\nabla^2$ and $R$ are the Laplacian and the curvature scalar in the metric $f_{\alpha\beta}$, and $\zeta$ is the conformal coupling.  The magnitude of $\zeta$ depends on the dimension of superspace and vanishes in the case of $2D$, which is of interest to us here. 
We see from Eq.(\ref{HG}) that $G$ is a Green's function of the WDW equation. 

For the KS model the Hamiltonian is quadratic and the path integral in (\ref{G}) can be performed exactly.  Then, up to an overall multiplicative constant, HL found that the amplitude reduces to 
\beq
G(b,c|b',c')=\int_0^\infty \frac{dN}{N}\exp\left[\frac{i}{2}\left(\alpha N-\frac{\beta}{N}\right)\right],
\label{GN}
\eeq
where 
\beq
\alpha=1-\frac{H^2}{3}(b^2+bb'+{b'}^2),~~~ \beta=(2\pi)^2 (c-c')(b-b').
\label{alphabeta}
\eeq
This amplitude should satisfy
\beq
{\cal H}G(b,c|b',c')\propto -i\delta(b-b')\delta(c-c').
\label{Green}
\eeq

The integral over $N$ in (\ref{GN}) can be expressed in terms of Bessel functions \cite{HL}.  With Nariai initial conditions (\ref{Nic}) we have
\beq
\alpha=-\frac{1}{3}(Hb-1)(Hb+2),~~~ \beta=\left(\frac{2\pi}{H^2}\right)^2 (H^3 c-1)(Hb-1)
\eeq
and 
\beq
G=-i\pi H_0^{(2)}\left(\sqrt{-\alpha\beta}\right)~~~~~(H^3 c>1)
\label{GH0}
\eeq
\beq
G=2 K_0\left(\sqrt{\alpha\beta}\right)~~~~~(H^3 c<1)
\label{GK0}
\eeq

The amplitude $G$ in Eqs.(\ref{GH0}),(\ref{GK0}) is similar to the IKTV wave function (\ref{PsiMI}),(\ref{PsiMI2}).  The difference is in the prefactor and in the index of the Bessel functions.  The two objects are closely related, as we will now show.

The Bessel functions appearing in Eqs.(\ref{GH0}),(\ref{GK0}) can be expressed as $Z_0(X)$, where
\beq
X=\sqrt{H^3 c-1}f(b)=4\pi{\phi}\sqrt{H^2{\tilde a}^2-1}
\eeq
with
\beq
f(b)=\frac{2\pi}{H^2}(Hb-1)\sqrt{\frac{Hb+2}{3}}=4\pi{\phi},
\eeq
where we have used the notation $Z_\nu$ for a Bessel function (of any kind) with index $\nu$ and Eqs.(\ref{tildeaphi}) relating $a$ and $b$ to ${\tilde a}$ and ${\phi}$.
Differentiating twice with respect to $c$, we obtain
\beq
\frac{\p^2 Z_0}{\p c^2}=\frac{H^6 f^2(b)}{4(H^3 c-1)}\left[-Z_1'+\frac{1}{f(b)\sqrt{H^3 c-1}}Z_1\right]=\frac{H^6 f^2(b)}{4(H^3 c-1)}Z_2 =(2\pi H^3)^2 \frac{{\phi}^2}{{\tilde a}^2-1}Z_2.
\label{Z}
\eeq
Here prime stands for a derivative with respect to the argument and we used an iteration formula for Bessel functions in the second step.

Now, the expression on the right-hand side of Eq.(\ref{Z}) has the same form as the wave function (\ref{PsiMI}) of IKTV.
We conclude that
\beq
\Psi_{IKTV}=C\frac{\p^2}{\p c^2}G(b,c|H^{-1},H^{-3})
\label{JTKS}
\eeq
with $C={\rm const}$.
Furthermore, it follows from Eq.(\ref{Green}) that\footnote{Note that $\p/\p c$ commutes with ${\cal H}$.}
\beq
{\cal H}\Psi_{IKTV}\propto-i\delta(b-H^{-1})\delta''(c-H^{-3}).
\eeq
Hence $\Psi_{IKTV}$ is not a solution of the WDW equation.  It has a distributional source at $a=b=H^{-1}$, which is more singular than that of a Green's function.

\section{Hartle-Hawking wave function}
 
In the original Hartle \& Hawling paper \cite{HH} the HH wave function is defined as 
\beq
\Psi(g_b) = \int^{g_b} {\cal D} g e^{-S_E(g)},
\label{PsiHH1}
\eeq 
where the integration is over "regular" $4D$ Euclidean geometries $g$, having a single boundary ${\cal B}$ with a 3-metric $g_b$.  For simplicity we specialize to models without any matter fields.     
As it stands, this definition is rather problematic.  The Euclidean action $S_E$ is unbounded from below, so the integral in (\ref{PsiHH1}) is divergent.  This can often be dealt with by a suitable analytic continuation of the integration variables.  Another problem is that the metrics contributing to the path integral are generally rather irregular, even non-differentiable.  So the notion of integrating over regular geometries needs to be defined.  The same problem arises in JT gravity.

IKTV attempted to get around this issue by focusing on the upper limit of integration in (\ref{PsiHH1}), with the hope that the regularity condition would somehow take care of itself.  They allowed the boundary curve of the $2D$ geometry to fluctuate and required that the asymptotic form of the wave function agrees with the semiclassical pre-exponential factor resulting from these fluctuations.  Another approach that they used was to calculate the path integral (\ref{PsiHH1}) for JT gravity with $\Lambda<0$, where it is better defined, and then analytically continue to $\Lambda>0$.  IKTV find that the two approaches agree and yield the wave function (\ref{PsiMI}).  Our analysis shows, however, that this wave function is unsatisfactory, as it is not a solution of the WDW equation.  Instead, it is related to the transition amplitude from a Nariai initial state with $a=H^{-1}$ and $\phi=0$.  The path integral (\ref{G}) for this amplitude is over geometries with two boundaries -- one with the specified values of $a$ and $\phi$ and the other with the `Nariai' values.  This does not square well with the intuitive idea of quantum creation of the universe from nothing.

We therefore need to revisit the question of how the path integral over regular Euclidean geometries has to be defined.  In the context of minisuperspace KS model, this issue has been discussed in detail by HL \cite{HL}, whose approach we largely follow.  

\subsection{Boundary conditions}

To discuss the boundary conditions for the no-boundary path integral, it is more convenient to represent the Euclideanized KS metric as 
\beq
ds_E^2 = N^2 dt^2+a^2(t)dx^2+b^2(t) d\Omega^2 .
\label{KSE}
\eeq
The Euclidean action is then \cite{HL} 
\beq
S_E= \pi\int_0^1 dt \left[-\frac{1}{N}\left(a{\dot b}^2+2b{\dot a}{\dot b}\right)+Na(H^2 b^2-1)\right]+S_b,
\label{S1}
\eeq
where $S_b$ is the boundary term\footnote{ The Gibbons-Hawking boundary terms in the gravitational action cancel out after integration by parts if the geometry has two boundaries -- at $t=0$ and $t=1$.  But for a compact geometry with a single boundary at $t=1$, the boundary term at $t=0$ has to be kept \cite{HL}.}
\beq
S_b=-\left[\frac{\pi}{N}\frac{d}{dt}(ab^2)\right]_{t=0}.
\label{Sb1}
\eeq
The time variable $t$ is defined so that $0<t<1$ with $t=1$ corresponding to the boundary ${\cal B}$ and $t=0$ corresponding to the "bottom" ${\cal B}_0$ of the 4-geometry $g$.  

The boundary conditions at $t=1$ fix the values of $\{a,b\}$, while the boundary conditions at $t=0$ should be chosen so that the geometry closes smoothly at ${\cal B}_0$.  HL show that for a classical 4-geometry(\ref{KSE}) there are two choices:
\beq
a(0)=0,~~\frac{1}{N}{\dot a}(0)=\pm 1, ~~ \frac{1}{N}{\dot b}(0)=0
\label{classbc1}
\eeq
and
\beq
b(0)=0,~~\frac{1}{N}{\dot b}(0)=\pm 1, ~~ \frac{1}{N}{\dot a}(0)=0,
\label{classbc2}
\eeq
where overdots now stand for derivatives with respect to $t$.

The time derivatives ${\dot a}$ and ${\dot b}$ are related to the (Euclidean) momenta conjugate to $a$ and $b$:
\beq
p_a=-\frac{2\pi}{N}b{\dot b},~~ p_b=-\frac{2\pi}{N}(a{\dot b}+b{\dot a}),
\eeq
so these boundary conditions correspond to fixing $\{a,p_a,p_b\}$ or $\{b,p_a,p_b\}$ at ${\cal B}_0$.  It is however inconsistent in quantum theory to fix a coordinate and its conjugate momentum.  Hence the best one can do is to impose  two out of the three conditions, for example
\beq
a(0)=0,~~p_b(0)=\mp 2\pi b(0)
\label{bc1}
\eeq
or
\beq
b(0)=0,~~p_a(0)=0.
\label{bc2}
\eeq
HL note that if classical field equations hold, then with either of these choices all three conditions in (\ref{classbc1}) or (\ref{classbc2}) are satisfied.  One can therefore expect that in the semiclassical regime the path integral will be dominated by regular geometries.
Since we are interested in dimensional reduction of KS model with the $S^2$ part integrated out, we will focus on the boundary conditions (\ref{bc1}), which correspond to fixing $a$ and ${\dot a}$ on ${\cal B}_0$.

HL suggest that a better choice of variables, suitable for the boundary conditions (\ref{bc1}), would be
\beq
A=2\pi b^2,~~ B=2\pi ab
\eeq
with the conjugate momenta
\beq
P_A=-\frac{\dot a}{2N},~~P_B=-\frac{\dot b}{N}.
\eeq
The boundary conditions (\ref{bc1}) then take the form
\beq
B'=0,~~{P_A}'=\mp \frac{1}{2},
\label{bc}
\eeq
where primes indicate the values at $t=0$.

The HH wave function can now be expressed as \cite{HL}
\beq
\Psi_{NB}(A,B) =G(A,B|{P_A}',B')=\int dN\int{\cal D}Q^\alpha{\cal D}P_\alpha\exp(-S_E),
\eeq
where the path integral is taken over histories with fixed $\{A,B\}$ and $\{{P_A}',B'\}$.  Unlike the case of fixed initial values $a'$ and $b'$,  this path integral cannot be evaluated exactly. { We therefore use the WKB method to express $\Psi_{HH}$ approximately as 
\beq
\Psi_{HH}(Q^{\alpha})=\int_C  \mu\left(Q^{\alpha},Z^{\beta},N\right)\exp\left[-S_{E}(Q^{\alpha};N|Z^{\beta})\right]dN,
\label{PsiZ}
\eeq
where $Q^\alpha=\{A,B\},~Z^\beta=\{{P_{A}}',B'\}$, $\mu$ is the semiclassical prefactor of the propagator and}
\beq
S_{E}=\pi\left[ \frac{H^2}{3}N(b^2+bb'+{b'}^2)-N-\frac{1}{N}(b-b')\left(a^2 b-\frac{{B'}^2}{b'}\right)+2b'^2{P_{A}}'+2B'{P_{B}}'\right]
\label{SE0}
\eeq
is the Euclidean action evaluated on a history satisfying the boundary conditions and the second order equations (\ref{1}),(\ref{2}) for $a$ and $b$, but not the constraint equation (\ref{3}).  {Note that the last term in (\ref{SE0}) can be neglected since it does not contribute to the path integral and the semiclassical prefactor.} The integration contour $C$ in (\ref{PsiZ}) is generally complex; we shall discuss the choice of this contour in Sec.4.4.   The calculation of the prefactor is discussed in Sec. 4.3.


HL discussed the calculation of the HH wave functions for KS model only for the case of a vanishing cosmological constant, $H=0$.  Eqs.(\ref{PsiZ}),(\ref{SE0}) apply for arbitrary $H$, but from this point on we cannot directly use the results of HL and will have to extend their analysis to $H>0$.

The initial value $b'$ in Eq.(\ref{SE0}) has to be expressed in terms of the boundary values $A,B$ (or $a,b$), $B',{P_{A}}'$, and the lapse $N$.  This can be done using the solutions ${\bar a}(\tau),{\bar b}(\tau)$ of the second order field equations (\ref{1}),(\ref{2}) (but not of the constraint equation).  HL give these solutions in terms of the time variable ${\tau}$, which is related to $t$ as $d{\tau}=a(t)dt$:
\beq
{\bar b}({\tau})=(b-b'){\tau}+b'
\label{eq1}
\eeq
\beq
{\bar a}^2({\tau}) {\bar b}({\tau})=-\frac{H^2 N^2}{3}(b-b'){\tau}^3-H^2 N^2 b'{\tau}^2 +\left[a^2b-{a'}^2 b'+\frac{H^2 N^2}{3}(b+2b')\right]{\tau} +{a'}^2 b'.
\label{eq2}
\eeq 
Expressed in terms of $\tau$, the boundary condition $(1/N)(da/dt)=\pm 1$ takes the form 	
\beq
\frac{\bar a}{N}\frac{d{\bar a}}{d{\tau}}({\tau}=0)=\mp 2{P_{A}}'=\pm 1.
\label{bctautilde}
\eeq
To implement this boundary condition, we differentiate Eq.(\ref{eq2}) with respect to ${\tau}$ and take the limit ${\tau}\to 0$. This gives (after dividing by ${b'}^2$)
\beq
\frac{4N{P_{A}}'}{b}{b'}=-a^2+\frac{{B'}^2}{{b'}^2} -\frac{H^2 N^2}{3b}(b+2b').
\label{b'eq}
\eeq
where we have used $a'=B'/b'$.

We now have to solve Eq.(\ref{b'eq}) for $b'$, substitute the result into the action (\ref{SE0}), and use it to evaluate $S_{E}$ and the pre-factor in (\ref{PsiZ}) with the values of $P_A'$ and $B'$ specified by the boundary conditions (\ref{bc}).  This calculation is significantly simplified if we note that the solution of Eq.(\ref{b'eq})
minimizes the action (with our boundary conditions), and thus $\p S_{E}/\p b'=0$.  It follows that when calculating the derivative of $S_{E}$ with respect to $B'$ in the determinantal prefactor in (\ref{PsiZ}), we only need to account for an explicit dependence of $S_{E}$ on this variable and can disregard the dependence through $b'$.  This means that we can substitute the boundary value $B'=0$ directly in Eq.(\ref{b'eq}).  Then, instead of a cubic equation we get a linear equation for $b'$, with the solution
\beq
b'=-\frac{b}{2N}\frac{a^2+H^2 N^2/3}{{2P_{A}}'+H^2 N/3}.
\label{b'b}
\eeq
Substituting this in the action (\ref{SE0}) we find
\beq
S_E=\frac{\pi N}{3}(H^2 b^2-3)-\frac{\pi}{N}a^2 b^2-\frac{\pi b^2}{4N^2}\frac{\left(a^2+\frac{H^2 N^2}{3}\right)^2}{2P_{A}'+\frac{H^2 N}{3}}-\frac{\pi{B'}^{2}}{N}\frac{\left(a^{2}+H^2 N^{2}+4NP_{A}^{\prime}\right)}{a^{2}+H^2 N^{2}/3}.
\label{SE1}
\eeq
where we have not substituted the boundary values ${P_{A}}^{\prime}$, $B^{\prime}$ yet, in order to calculate the prefactor.

\subsection{Saddle points}

Without making any approximations for the action, the saddle points cannot be found in closed form.  Since we will be mostly interested in the regime where $b\approx H^{-1}$, we will first find the saddles for $b=H^{-1}$ and then treat deviations from these points as small perturbations.

We also have to decide on the choice of sign in the boundary condition (\ref{bc}) for $P_A'$. {Here we will follow HL and pick $P_A'=-1/2$.  Their justification is that for this choice the final boundary $\mathcal{B}$ is to the "future" of the initial boundary $\mathcal{B}_{0}$. In fact, there is a stronger argument: 
it can be shown that choosing the opposite sign in (\ref{bc}) does not yield convergent contours for the HH wave function. Furthermore, the characteristic exponential factor $\exp(\pi/H^{2})$ can only be retrieved with the choice $P_A'=-1/2$. }


Thus, setting $b=H^{-1}$, $B'=0$ and $P_A'=-1/2$ we have
\begin{equation}
S_{E0}=-\frac{2\pi N}{3}-\frac{\pi a^{2}}{H^2 N}-\frac{\pi\left(3a^{2}+H^2 N^{2}\right)^{2}}{12H^2 N^{2}\left(H^2 N-3\right)},
\label{SEN}
\end{equation}
 where the extra subscript "0" indicates that the action is evaluated at $Hb=1$.
Note the singularities at $N=0$ and $N=3/H^2$. 

For the following analysis it will be convenient to introduce the rescaled variables 
\beq
u=H^2 a^{2} \ \ , \ \ \tilde{N}=H^2 N \ \ , \ \ \tilde{S_{E}}= \frac{H^2 S_{E}}{\pi}
\eeq
The rescaled action (\ref{SEN}) is given by
\beq
{\tilde S}_{E0}=-\frac{2\tilde{N}}{3}-\frac{u}{ \tilde{N}}-\frac{\left(3u+\tilde{N}^{2}\right)^{2}}{12\tilde{N}^{2}\left(\tilde{N}-3\right)}
\eeq

In order to evaluate the integral (\ref{PsiZ}) using the method of steepest descent, we  first determine the 
extrema of the action $\tilde{S}_{E0}$. These are given by
\beq
\frac{\p \tilde{S_{E0}}}{\p \tilde{N}}=0.
\eeq
The resulting equation is quintic in $\tilde{N}$:
\beq
\left(\tilde{N}^{2}-2\tilde{N}+u\right)\left(\tilde{N}^{3}-4\tilde{N}^{2}-3\tilde{N}u+6u\right)=0
\label{SSol}
\eeq
Its solutions are
\beq
\tilde{N}_{1,2}=1\pm\sqrt{1-u}
\label{N12}
\eeq
\beq
\tilde{N}_{3}=\frac{4}{3}+\frac{16+9u}{3A}+\frac{A}{3}
\eeq
\beq
\tilde{N}_{4,5}=\frac{4}{3}-\frac{\left(1\mp i\sqrt{3}\right)(16+9u)}{6A}-\frac{\left(1\pm i\sqrt{3}\right)A}{6}
\eeq
where the quantity $A$ is given by
\beq
A=(64- 27 u + 
9\sqrt{-128 u-39u^{2} -9u^{3}})^{1/3}
\eeq

The solutions $3,4,5$ are expressed here in a rather complicated form. Taking a closer look at the quantity $A$ we can express it in a more convenient way with the Euler representation of complex numbers. After some straightforward calculations we arrive at
\beq
A=\sqrt{16+9u} \ e^{\textstyle i\frac{\theta}{3}},
\eeq
where $\theta$ is given by
\beq
\theta=\arctan\left[\frac{9\sqrt{9u^{3}+39u^{2}+128u}}{64-27u}\right]
\label{theta1}
\eeq
The saddles $\tilde{N}_{3,4,5}$ are then simplified to:

\beq
\tilde{N}_{3}=\frac{4}{3}+\frac{2\sqrt{16+9u}}{3}\cos{\frac{\theta}{3}}
\eeq

\beq
\tilde{N}_{4,5}=\frac{4}{3}-\frac{2\sqrt{16+9u}}{3}\cos\left[{\frac{\theta\pm\pi}{3}}\right]
\label{N45}
\eeq
It is clear that the saddles $3,4,5$ are always real.  ${\tilde N}_1$ and ${\tilde N}_2$ are also real for $u\leq 1$, while for $u>1$ they form a complex conjugate pair.

\subsection{Prefactor}
{The semiclassical prefactor for the propagator is given by \cite{Schulman} 
\beq
\mu=f^{-1/4}\sqrt{D}f'^{-1/4}
\label{mu1}
\eeq
where $f^{\prime}$ and $f$ are the determinants of the minisuperspace metric $f_{\mu\nu}$ evaluated at $t=0$ and $t=1$ respectively
and $D$ is the Van Vleck-Morette determinant \cite{VV}\cite{Morette}. In the representation $A=2\pi b^{2}$ and $B=2\pi ab$ the Hamiltonian for the KS model takes the form \cite{HL}
\beq
\mathcal{H}=-P_{B}^{2}-\frac{4AP_{A}P_{B}}{B}+1-\frac{H^{2}}{2\pi}A=\frac{1}{2}f^{\mu\nu}P_{\mu}P_{\nu}+V
\eeq
where $V=1-A H^{2}/(2\pi) \ $. Thus the minisuperspace metric and its determinant are
\beq
f_{\mu\nu}=
\begin{pmatrix}
\frac{B^{2}}{8A^{2}} & -\frac{B}{4A}\\
-\frac{B}{4A} & 0
\end{pmatrix},
~~~~f\propto\frac{B^{2}}{A^{2}}
\eeq
\newline
{The action (\ref{SE1}) expressed in variables $\{A,B\}$ is given by
\beq
S_{E}=\frac{\pi N}{3}\left(\frac{AH^{2}}{2\pi}-3\right)-\frac{B^{2}}{4N\pi}-\frac{A}{8N^{2}}\frac{\left(\frac{B^{2}}{2\pi A}+\frac{H^{2}N^{2}}{3}\right)^{2}}{2P^{\prime}_{A}+\frac{H^{2}N}{3}}-\frac{{\pi B^{\prime}}^{2}}{N}\left(\frac{\frac{B^{2}}{2\pi A}+H^{2}N^{2}+4NP^{\prime}_{A}}{\frac{B^{2}}{2\pi A}+\frac{H^{2}N^{2}}{3}}\right)
\eeq
and the Van Vleck-Morette determinant $D$ can be calculated as
\beq
D\equiv\det\left[-\frac{\p^2 S_E}{\p Q^\alpha \p Z^\beta}\right]=\frac{3B B^{\prime}}{A N^{2}\left(H^{2}N+6P^{\prime}_{A}\right)},
\eeq
where, as before, $Q^\alpha=\{A,B\},~Z^\beta=\{{P_{A}}',B'\}$. Inserting the above relations in Eq.(\ref{mu1}) for the prefactor and switching back to variables $\{a,b\}$,} we obtain\footnote{We note that there is an error in the expression (6.5) for the prefactor in Ref.\cite{HL}.  We are grateful to Jorma Louko for a discussion of this point.}
\beq
\mu\left(a,b,N\right)\propto \frac{b^{\prime}}{N\sqrt{H^{2}N-3}}.
\label{mu}
\eeq
where $b^{\prime}$ is a function of $a,b,N$ and is given by Eq.(\ref{b'b})  and we have now inserted the boundary value of $P^{\prime}_{A}=-1/2$.

The prefactor in Eq.(\ref{mu}) introduces a branch cut, which we can choose to lie at $N> 3/H^{2}$ along the real axis.
From the analysis that follows, we will see that the choice of a suitable contour will not be affected significantly by this branch cut.


\subsection{Integration contours}

One of the key issues that remains to be addressed is the choice of the integration contour over $N$ in Eq.(\ref{PsiZ}).  
No general prescription for this choice has yet been given.  Integration over real or pure imaginary values of $N$ yields divergent integrals, so one has to look for a non-trivial contour in the complex plane that would make the integral convergent.  The consensus view appears to be
that the contour $C$ should satisfy the following three criteria (see, e.g., \cite{Halliwell:1989dy}).  (1) $C$ should not have ends: it must be either infinite or closed.  This guarantees that $\Psi_{HH}$ is a solution of the WDW equation (rather than a Green's function).  (2) The HH wave function should be real.  This can be achieved by choosing a contour which is symmetric with respect to the real $N$ axis.  This requirement can be thought of as an expression of the $CPT$ invariance of the HH state \cite{HHH}.  (3) The wave function should predict a classical spacetime when the universe is large.  This means that in the appropriate limit $\Psi_{HH}$ should be a superposition of rapidly oscillating terms of the form $e^{iS}$, where $S$ is the classical action.  We shall adopt these criteria as the defining properties of $\Psi_{HH}$.
\begin{figure}[b!]
	\centering
		\includegraphics[scale=0.7]{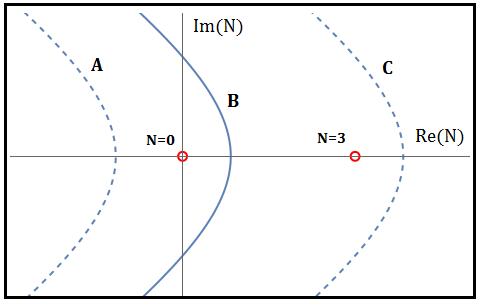}
		\caption{Examples of convergent infinite contours in the complex $N$ plane. The singularities $N=\{0,3\}$ are shown in circles.}
		\label{fig1}
	
\end{figure}

Let us first consider infinite contours.  From Eq.(\ref{SEN}) we find that $S_E\sim -3N/8$ for $|N|\to\infty$.  It follows that the integral over $N$ can be convergent only if the asymptotic directions of the contour are at $|{\rm arg}~N|>\pi/2$.  Some inequivalent choices are illustrated in Fig.\ref{fig1}.   These contours are symmetric with respect to the real axis, so they define a real wave function.  We will first consider the contour $B$ which crosses the real axis at $0<N<3$ and will comment on the other choices at the end of this section.  Note that the contour $B$ may almost coincide with the imaginary axis.  It could cross the real axis at $N=+\epsilon$ and asymptote to ${\rm arg}~N=\pm (\pi/2+ \epsilon')$, where $\epsilon,\epsilon' \to +0$. 

The integration contour $B$ can be turned into a closed contour by adding to it an infinite arc $|N|={\rm const}\to\infty$.  The integral over the arc vanishes in the limit, so the original integral remains unchanged.
The resulting contour can be distorted into a finite closed contour which encircles the singularity at $N=0$ but does not encircle the singularity at $N=3$.  Thus the infinite contours of type $B$ are equivalent to this kind of closed contours. 
\begin{figure}[b!]
	\centering
		\includegraphics[scale=0.5]{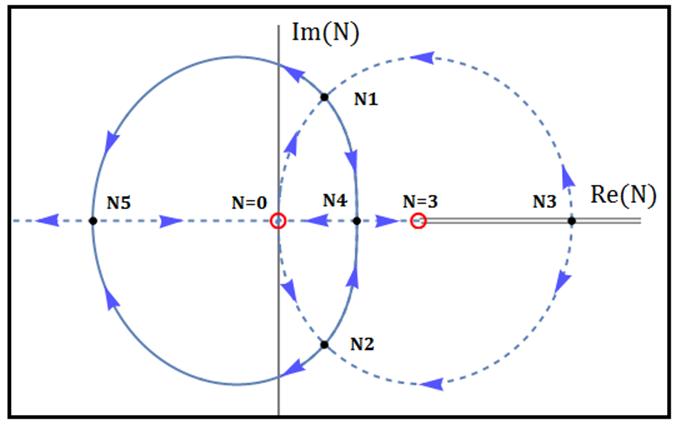}
		\caption{The steepest descent contours for $u>1$  and $Hb=1$.
		The arrowheads point to the direction where $Re(-\tilde{S}_{E})$ decreases. The saddles $\tilde{N}_{i}$ are marked with solid dots and the singularities with circles. Note the branch cut at $\tilde{N}\in(3^{+},+\infty)$. The HH contour corresponds to the solid curve encircling the singularity $N=0$; it is dominated by saddles $N_{1}$, $N_{2}$.}
		\label{fig2}
	
\end{figure}
Following the Picard-Lefschetz prescription,\footnote{ For a simple review of Picard-Lefschetz theory see, e.g., Ref.~\cite{Feldbrugge:2017kzv}.} the closed contour can now be distorted so that it passes through saddle points following the steepest descent and ascent lines, making the integral absolutely convergent.  Let us first consider the case of $a>H^{-1}$.  The saddle points for this case are shown in Fig.\ref{fig2}.  The steepest descent and ascent lines are defined by ${\rm Im}~S_E={\rm const}$.  The lines passing through the saddle points are also shown in the figure, with arrows indicating the direction in which $-{\rm Re}~S_E$ is decreasing.  The contour encircling the singularity at $N=0$ can be distorted into the contour passing through the saddle points $N_1$, $N_4$, $N_2$ and $N_5$.  This contour is dominated by the saddles at $N_1$ and $N_2$.

We now consider the case of $a<H^{-1}$, when all saddle points lie on the real axis.  The steepest descent and ascent lines in this case are illustrated in Fig.\ref{fig3}. The contour encircling $N=0$ can now be deformed into the contour passing through $N_4$ and $N_5$.  It is dominated by the saddle at $N_4$.

\begin{figure}[h!]
	\centering		\includegraphics[scale=0.59]{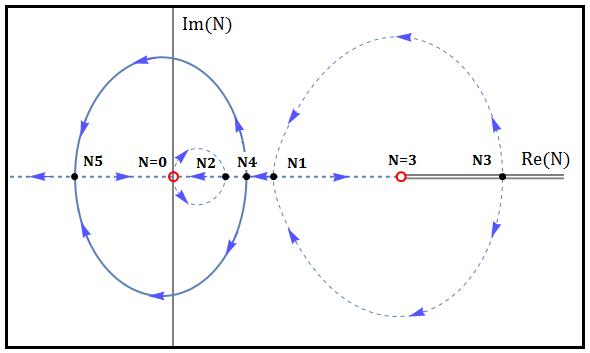}
		\caption{	
		{The steepest descent contours for $u<1$  and $Hb=1$. In this case, all the saddles are real.
		 The HH contour corresponds to the solid curve encircling the singularity $N=0$ and dominated by saddle $N_{4}$.} 
		 }
		\label{fig3}
\end{figure}

{We finally comment on other possible choices of the integration contour.}
A contour of type $A$ in Fig.1 crosses the real axis at $N<0$.  After it is closed by adding an infinite arc, this contour does not encircle any singularities, so it can be continuously shrunk to a point.  The wave function defined by this contour is therefore identically zero.

{Another possibility is to choose the branch cut to lie at ${\tilde N}<3$ on the real axis and choose the contour that crosses the real axis at ${\tilde N}>3$ (see contour $C$ in Fig.1).
For $a > H^{-1}$ such a contour can be deformed into a contour that runs along the steepest descent/ascent lines from $N_3$ to $N_1$, then takes a turn and goes to $N_5$, and from there runs above the branch cut along the real axis to $N\to -\infty$.  This has to be supplemented by another half of the contour that runs symmetrically from $N_3$ through $N_2$ and $N_5$ to $-\infty$ in the lower half-plane.}
The resulting wave function is then non-oscillating, with the main contribution given by the {real} saddle point $N_3$.  This is in conflict with the defining property (3) of the HH wave function.  We thus conclude that the only acceptable choice of integration contour is an infinite contour of type B or equivalently a closed contour encircling the singularity at $N=0$.

\subsection{Perturbing the Saddle Points}

To make a connection with JT gravity, we need to know the KS wave function for $b$ very close but not equal to $H^{-1}$.
The saddle points and the steepest descent/ascent lines will then be slightly different from those we found in the subsections B and D.  For $a>H^{-1}$ we are interested in the complex saddle points $N_{1,2}$ which dominate the integral.  Let us define the shift $x$ of the saddle point $\tilde{N}_{i}$ as
\beq
\tilde{N}=\tilde{N}_{i}+x ,
\label{Spert}
\eeq
where $i=1,2$, $\tilde{N}_{i}$ are given by (\ref{N12}) and $|x|<<1$. We insert this into the action and expand to second order in $x$.  This gives :
\beq
\tilde{S}_{E}\approx-1\mp 2i (1-Hb)\sqrt{u-1}-2(1-Hb)x+f(u)x^{2}+O(x^{3})
\label{expansion}
\eeq
where the upper and lower signs are for $N_1$ and $N_2$ respectively and {the function $f(u)$ is given in the Appendix.}

The action is extremized with
\beq
x=\frac{1-Hb}{f(u)}.
\label{x}
\eeq
Since $x$ depends linearly on $(1-Hb)$ the contribution of the x-dependent terms to the action is ${\cal O}[(1-Hb)^{2}]$.  Thus the action of the dominant saddle points is
\beq
S_E\approx -\frac{\pi}{H^2}\mp\frac{2i\pi}{H^2}(1-Hb)\sqrt{H^2a^2-1}+{\cal O}[(1-Hb)^2].
\label{SEab}
\eeq

The steepest descent/ascent lines can be calculated numerically for any values of $a$ and $b$.  For $Ha>1$ the character of these lines changes when $Hb$ is moved away from 1, even for an arbitrarily small amount.  For $Hb>1$, the steepest descent contour passing through $N_1$ follows nearly the same path as for $Hb=1$, but short of reaching $N_5$ it makes a turn and runs along the real axis towards $N=0$.  Then it runs back, repeats the same path symmetrically in the lower half-plane and arrives at $N_2$.  From there it runs towards $N_4$ following nearly the same path as for $Hb=1$, but short of reaching $N_4$ it makes a turn and runs towards the singularity at $N=3$.  Finally it runs back symmetrically and returns to $N_1$.  This contour is illustrated in Fig.\ref{pert1a} .  The only change compared to the original $Hb=1$ contour is that small segments near $N_4$ and $N_5$ are now replaced by sharp spikes running towards the singularities and back.  The integrals over the upper and lower halves of these spikes nearly cancel one another, so their combined contribution to the wave function is very small.
\begin{figure}[t!]
	\centering		\includegraphics[scale=0.66]{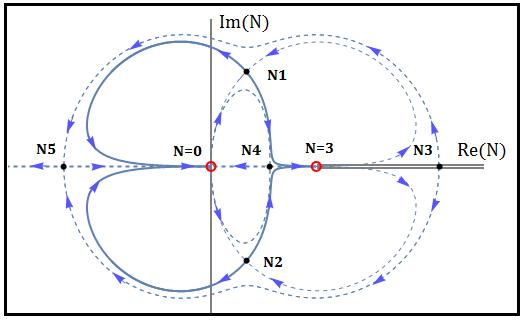}
		\caption{{The perturbed steepest descent contours for $u>1$ and $Hb>1$. The HH contour does not pass through the saddles $N_{4}$ and $N_{5}$.}}
		\label{pert1a}
\end{figure}
\begin{figure}[h!]
	\centering		\includegraphics[scale=0.74]{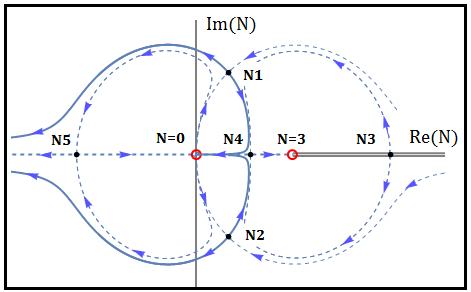}
		\caption{{The perturbed steepest descent contours for $u>1$ and $Hb<1$. The HH contour does not pass through the saddles $N_{4}$ and $N_{5}$.}}. 
		\label{pert1b}
\end{figure}

For $Hb<1$ the situation is very similar, except now instead of shooting to the right the spikes shoot to the left.  The spike originating near $N_5$ runs along the negative real axis to $-\infty$ and back, and the spike originating near $N_4$ runs to $N=0$ and back (see Fig.\ref{pert1b} ).  As before, the contour is dominated by the saddles $N_1$ and $N_2$, with the spikes making a very small contribution. 

 For $Ha<1$, small deviations of $Hb$ from 1 do not change the character of the steepest descent lines.  The contour is still dominated by the real saddle $N_4$, and the steepest descent line passing through this saddle also passes through $N_5$.  At some critical value of $Hb>1$ 
the saddles $N_1$ and $N_4$ merge, and at greater values they become a complex conjugate pair . A similar situation occurs for saddles $N_2$ and $N_4$ when $Hb<1$ . 
Generally, the saddle $N_{4}$ remains real in a range of $(Hb-1)$ that depends on $u$.  It can be shown that as $u$ increases from $0$ to $1$, the range of $(Hb-1)$ for which $N_{4}$ remain real shrinks from $\sim 1$ until it reaches zero at $u=1$, where the saddles $N_{1,2,4}$ merge.  For example, when $u\ll1$ the contour behavior does not change qualitatively for $0<Hb<\sqrt{3}$ and for $u=0.9$ the range is $|1-Hb|\sim0.008$.
It would be interesting to map the behavior of the contours and saddles for the full range of $a$ and $b$,
but we will not attempt this here.

\subsection{The Hartle-Hawking wave function}

We are now ready to calculate the semiclassical Hartle-Hawking wave function.

\subsubsection{$Ha>1$}

For $Ha>1$ we need to expand the action (\ref{SEN}) up to quadratic order in $(N-N_i)$ at saddle points $N_i$ ($i=1,2$) and then do the Gaussian integrals.  Up to an overall numerical factor, the corresponding contributions to the wave function are 
\beq
\frac{b^{\prime}(N_{i})}{N_i \sqrt{3-H^{2}N_i}}\sqrt{\frac{1}{S_{NN}(N_i)}}e^{-S_E(N_i)}.
\label{Psii}
\eeq
Here the factor $b^{\prime}(N_{i})/\left(N_i\sqrt{3-H^{2}N_i}\right)$ comes from Eq.(\ref{mu}), $S_E(N_i)$ from Eq.(\ref{SEab}) and 
\beq
S_{NN}=\frac{\p^2 S_E}{(\p N)^2}.
\eeq
At $N=N_i$ we have
\beq
S_{NN}(N_{1,2})=\frac{6 \pi\left(H^{2}a^{2}-1\right)}{a^{4}H^{2}\left(H^{2}a^{2}+3\right)}\left[-2\pm2i\sqrt{H^{2}a^{2}-1}\pm i H^{2}a^{2}\sqrt{H^{2}a^{2}-1}\right]
\eeq
and
\beq
b^{\prime}(N_{i})\left[N_i^2\left(3-H^{2}N_{i}\right) S_{NN}(N_i)\right]^{-1/2}=\frac{H}{\sqrt{6\pi\left(H^{2}a^{2}-1\right)}},
\eeq
Combining the contributions of the two saddle points, we obtain an approximate semiclassical HH wave function.  Up to an overall constant factor it is given by
\beq
\Psi_{HH}(Ha>1)\propto {\cal A}\exp\left(\frac{\pi}{H^2}\right)\cos\left(\frac{2\pi}{H^2}(1-Hb)\sqrt{H^2a^2-1}\right),
\label{PsiHHsemiclassical}
\eeq
where
\beq
{\cal A}=\frac{1}{\sqrt{H^{2}a^{2}-1}}.
\eeq 
{Note that we neglected corrections ${\cal O}(1-Hb)$ in the prefactor, but kept them in the exponent, which includes a large factor $H^{-2}$.} The WKB approximation is essentially an expansion in powers of $H$.
It is easily verified that the wave function (\ref{PsiHHsemiclassical}) satisfies the WDW equation to the leading order in $H$ and $(Hb-1)$.

As one might expect, $\Psi_{HH}$ exhibits the characteristic WKB divergence at the turning point $a=1/H$. This divergence is much milder than that in the IKTV solution.
The WKB approximation breaks down near the turning point, and we expect the exact wave function to remain finite there.

\subsubsection{$Ha < 1$}

For $Ha<1$ the integral over $N$ is dominated by the saddle point $N_4$.  This case is difficult to handle analytically, so we will only consider the limiting case  $Ha\ll 1$.  In this regime, we are able to go beyond the approximation $Hb\approx1$, as long as the qualitative behavior of the contours and the respective saddles does not change (see section 4.5).  

Let us first note that for $Ha \ll 1$ and $Hb \approx 1$, Eq.(\ref{N45}) gives 
\beq
H^2 N_4\approx \sqrt{\frac{3}{2}}Ha.
\label{N4appr}
\eeq
Let us further assume (to be verified shortly) that $H^2 {N}_4={\cal O}(Ha)$ in a wide range of $Hb\lesssim 1$. Then the action can be approximated by\footnote{ This can be seen from Eq.(\ref{Ssplit}) for $S_E$ in the Appendix by noticing that the first two terms in both of the big parentheses in that equation are ${\cal O}(Ha)$ while the last terms are ${\cal O}(H^2 a^2)$.}
\beq
S_{E}\approx\frac{\pi N}{3}(H^{2}b^{2}-3)-\frac{\pi}{N}a^{2}b^{2}
\eeq
and the corresponding saddles
 are
\beq
N_{4,5}=\pm\frac{ab}{\sqrt{1-H^{2}b^{2}/3}} .
\label{N4exactb}
\eeq
 We note that Eq.(\ref{N4exactb}) agrees with our assumption that $H^2 {N}_4={\cal O}(Ha)$ for $Hb\lesssim 1$, verifying that this assumption is consistent.
Substituting (\ref{N4exactb}) into the action we obtain
\beq
S_{E}\approx-\frac{2\pi ab}{\sqrt{3}}\sqrt{3-H^{2}b^{2}}.
\eeq
The WKB prefactor in the regime $Ha\ll 1$ can be found along the same lines as for $Ha>1$.  To lowest order in $Ha$, it is proportional to $\sqrt{a}$.
Thus, the wave function is given by
\beq
\Psi_{HH}\propto \sqrt{a}\exp\left(\frac{2\pi ab}{\sqrt{3}}\sqrt{3-H^{2}b^{2}}\right).
\label{PsiHa<1}
\eeq
This approximation applies for $H\ll Ha \ll 1$, which is a wide range in the sub-Planckian regime $H\ll 1$. Also, we have to constrain the values of $b$ to $Hb<\sqrt{3}$  in order for the saddles to remain real and the wavefunction to be non-oscillatory.  (Note that our approximation breaks down at $Hb\approx \sqrt{3}$, where the lapse parameter $N$ in Eq.(\ref{N4exactb}) becomes large.)

Keeping the scale factor $a$ fixed, the solution has a maximum at $Hb=\sqrt{3/2}$. This peak is not in the range $Hb\approx1$  which is of most interest for dimensional reduction to JT.  Note, however, that the maximum of the wave function (\ref{PsiHa<1}) at a fixed ${\tilde a}=a\sqrt{Hb}$ is at $Hb=1$.  This is more relevant, since ${\tilde a}$ plays the role of the scale factor after dimensional reduction.
A numerical WKB solution for $\Psi_{HH}$ in the full range $a<1/H$ with $Hb=1$ is shown in Fig.\ref{fig4}. 

Finally, it can be verified that the wave function (\ref{PsiHa<1}) satisfies the WDW equation to the leading order in $H$ and that it grows exponentially with $a$, as expected. 

\begin{figure}[h!]
	\centering
		\includegraphics[scale=0.6]{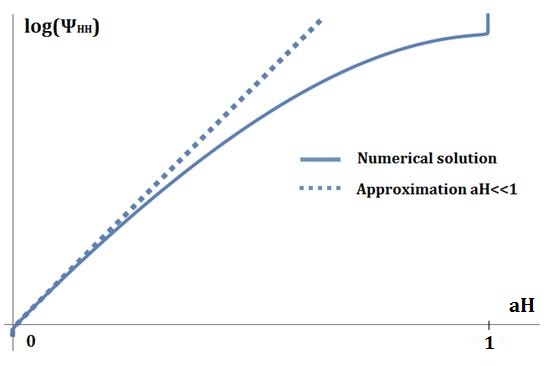}
		\caption{{A graph of $(\log\Psi_{HH},a)$ for $bH=1$.  A numerical WKB solution for the HH wavefunction is shown by the solid curve. It diverges abruptly at $aH=1$, due to the WKB prefactor.  The blue dashed line corresponds to the $aH<<1$ approximation. Note that both curves diverge to $-\infty$ at $a=0$ due to the pre-exponential factor $\sqrt{a}$ in (\ref{PsiHa<1}).}     }
		\label{fig4}
	
\end{figure}

\section{Probability distribution}

Probability distributions in minisuperspace quantum cosmology can be expressed in terms of the conserved current density
\beq
j^\alpha =i \sqrt{-f} f^{\alpha\beta}\left(\Psi^* \p_\beta\Psi-\Psi\p_\beta\Psi^*\right),
\label{jalpha}
\eeq
\beq
\p_\alpha j^\alpha=0,
\eeq
where $f^{\alpha\beta}$ is the minisuperspace metric and $f=\det (f_{\alpha\beta})$. 
In a $d$-dimensional minisuperspace one of the coordinates (or a combination of coordinates), call it $T$, can be designated as a "clock".  Then the probability distribution for the other coordinates on surfaces of constant $T$ is given by
\beq
dP\propto j^\alpha d\Sigma_\alpha,
\label{dPdSigma}
\eeq
where $d\Sigma_\alpha$ is the $(d-1)$-dimensional surface element.  
If the clock variable $T$ exhibits semiclassical behavior and its classical evolution is monotonic (which are the properties one can reasonably require from a good clock), then it can be shown that the probability defined by Eq.(\ref{dPdSigma}) is positive definite \cite{Vilenkin:1988yd}.

An obvious problem with the HH wave function is that it is real, so the current is identically zero.
 In the classically allowed range $Ha>1$ this can be circumvented if we calculate the current using only the branch of the wave function describing expanding universes.  Eq.(\ref{PsiHHsemiclassical}) for $\Psi_{HH}$ would then be replaced by
\beq
\Psi_{+}(Ha>1)\propto \frac{1}{\sqrt{H^2 a^2-1}} \exp\left[\frac{2\pi i}{H^2}(Hb-1)\sqrt{H^2a^2-1} +F(a)(Hb-1)^2 +{\cal O}\left((Hb-1)^3\right)\right],
\label{Psi+}
\eeq
 where we have dropped the constant factor $\exp\left[\pi/H^{2}\right]$.  We have also included a quadratic in $(Hb-1)$ correction in the exponential; we shall see that it plays an important role in the probability distribution.  The coefficient function $F(a)$ is given in the Appendix.

In the KS model, a natural choice for the clock variable is the scale factor $a$ (on the expanding branch of the wave function).  Alternatively, we can use $c=a^2 b$, since we are working in the regime where $b\approx H^{-1}={\rm const}$.  We shall adopt the latter choice, which is more convenient.  It is also more appropriate for the dimensional reduction, since $c\propto {\tilde a}^2$, where ${\tilde a}$ is the scale factor of the JT model. Then the probability distribution for $b$, Eq.(\ref{dPdSigma}), takes the form\footnote{Note that with $b$ and $c$ used as minisuperspace coordinates, the metric is $f^{bc}={\rm const}$, $f^{bb}=f^{cc}=0$, and its determinant is $f= -(f^{bc})^{-2} ={\rm const}$}
\beq
dP\propto j^c db,
\eeq
where
\beq
j^c\propto -i\left(\Psi^* \p_b\Psi-\Psi\p_b\Psi^*\right).
\eeq
Substituting the wave function $\Psi_+$ in $j^c$ and setting $Hb=1$ in the pre-exponential factor, we obtain
\beq
dP\propto \frac{db}{\sqrt{H^2 a^2-1}}\exp\left[2(Hb-1)^2 {\rm Re} F(a)\right].
\eeq

The real part of $F(a)$ is calculated in the Appendix:
\beq
{\rm Re} F(a)=-\frac{2\pi}{3H^2(H^2 a^2-1)}.
\eeq
Thus the probability distribution is
\beq
dP\propto  \frac{db}{\sqrt{H^2 a^2-1}}\exp\left[-\frac{4\pi(Hb-1)^2}{3H^2(H^2 a^2-1)}\right].
\label{P}
\eeq
There are a few interesting things to note about this distribution.  It is peaked at $b=H^{-1}$ with variance $\delta b\sim Ha$.  Our approximations are accurate for $\delta b\ll H^{-1}$, that is for 
\beq
H^{-1}\lesssim a\ll H^{-2}.
\label{acond}
\eeq 
The distribution is obviously normalizable.  Moreover, we note that
\beq
\int_{-\infty}^\infty\frac{dP}{db}db={\rm const},
\eeq
independent of $a$, so we can normalize the distribution to one.  This is a direct consequence of conservation of $j^\alpha$.

\section{Back to JT}

Once we have found the HH wave function $\Psi_{HH}^{(KS)}(a,b)$ for the KS model, we can {\it define} the HH wave function for JT gravity as the wave function obtained from $\Psi_{HH}^{(KS)}$ by dimensional reduction.  This amounts to expressing the scale factors $a,b$ in terms of the JT variables ${\tilde a},\phi$ using the connection formulas (\ref{tildeaphi}) and adding an extra factor of ${\tilde a}$ to account for the difference between $\Psi$ and $\tilde\Psi$ in the JT model. 

 An issue that needs to be addressed is that of the boundary conditions (\ref{bc}) that we used in the path integral for $\Psi_{HH}^{(KS)}$.  These boundary conditions were imposed to ensure a smooth closure of the geometry at $t= 0$; they are equivalent to $a(0)=0,~{\dot a}(0)/N=1$.  However, after dimensional reduction the new scale factor is given by ${\tilde a}=a\sqrt{Hb}$, so the appropriate boundary conditions would now be
\beq
{\tilde a}(0)=0,~{\dot {\tilde a}}(0)/N=1.
\label{abc}
\eeq
This is equivalent to (\ref{bc}) only if $Hb(0)=1$.
Requiring in addition that the dilaton is smooth at $t=0$, we should add the condition 
\beq
{\dot\phi}(0)=0.
\label{phibc}
\eeq

 As we discussed in Sec.4.1, a smooth closure of a classical $S^1\times S^2$ geometry requires three boundary conditions, while only two conditions can be consistently imposed in quantum theory.   Classically, the three conditions are not independent and any two of them imply the third.  In the present case the situation is similar: it follows from the boundary conditions (\ref{bc}) that $Hb(0)=1$ if the constraint equation (\ref{3}) is satisfied.  Thus the boundary conditions that we used are equivalent to smooth closure conditions for JT model at the classical level.  Quantum mechanically, different choices of two conditions out of three may be inequivalent, yielding wave functions satisfying WDW equations with different factor orderings.  One can expect, however, that in the semiclassical regime these wave functions will be close to one another, differing perhaps only in the pre-exponential factor.\footnote{ The semiclassical wave function for an FRW universe with a uniform scalar field $\phi$ was studied in Ref.\cite{Vilenkin:1987kf} for different factor orderings in the WDW equation.  A change of factor ordering had an effect on pre-exponential factor, but it did not affect the semiclassical probability distribution for $\phi$.  One can expect a similar situation to occur in the JT model.    
Since no particular choice appears to be preferred, 
we shall proceed to use our $\Psi_{HH}^{(KS)}$ for dimensional reduction.}



 For $Hb\approx 1$ the relations (\ref{tildeaphi}) become
\beq
Hb\approx1+2\phi H^{2} \ \ , \ \ a\approx\tilde a(1-\phi H^{2}) .
\eeq
Then in the classically allowed range $Ha>1$  the wave function (\ref{PsiHHsemiclassical}) becomes 
\beq 
\Psi_{HH}\sim \frac{\tilde{a}}{\sqrt{H^2 \tilde{a}^2-1}}\exp\left(\frac{\pi}{H^2}\right)\cos\left(4\pi\phi\sqrt{H^2 \tilde{a}^2-1}\right) ,
\label{HHJT}
\eeq
were we have neglected $H^2$ corrections to the prefactor. 


In the classically forbidden region we can use the small-$Ha$ solution (\ref{PsiHa<1}) to obtain the JT wave function for $H{\tilde a}\ll 1$.  We find\footnote{For this calculation, it is helpful to rearrange the relation of $\phi$ and $b$ in the form:$\sqrt{\frac{2}{Hb}}\sqrt{1-6\phi^{2}H^{4}}=\sqrt{3-H^{2}b^{2}}$.}
\beq
\Psi_{HH} \sim \tilde{a}^{3/2} \exp\left[\frac{4\tilde{a}\pi}{\sqrt{6}H}\sqrt{1-6\phi^{2}H^{4}}\right] ,
\label{HHJT<1}
\eeq
where we have neglected corrections to the prefactor. This expression is valid for $\tilde{a}H\ll 1$ and $\phi^{2}H^{4}<1/6$. The wave function peaks at $\phi=0$ at a fixed $\tilde{a}$. It can be shown that it satisfies the WDW equation of JT gravity (\ref{WDW5}) to the leading order.

Similarly to the KS model, the expanding branch of the wave function (\ref{HHJT}) can be used to find the probability distribution for the dilaton field $\phi$. The difference here is that now the conserved current is given by Eq.(\ref{jalpha}) with $\Psi$ replaced by ${\tilde\Psi}=\Psi/a$.  The reason is that the differential operator in Eq.(\ref{WDW5}) is not the covariant Laplacian, because of nonstandard factor ordering in Henneaux's quantization of the JT model.
As a result the probability distribution is obtained by simply using the connection formulas (\ref{tildeaphi}) in Eq.(\ref{P}) without any extra factors of ${\tilde a}$:
\beq
dP\propto  \frac{d\phi}{\sqrt{H^2 {\tilde a}^2-1}}\exp\left[-\frac{16\pi H^2\phi^2}{3(H^2 {\tilde a}^2-1)}\right].
\label{Pphi}
\eeq
We expect this expression to be accurate for $\phi\lesssim H^{-1}\delta b\ll H^{-2}$ and ${\tilde a}$ satisfying the conditions (\ref{acond}).

We note that the classical solutions (\ref{solution}) of the JT model
\beq
a=H^{-1} \cosh (Ht), ~~~~ \phi=\phi_0 \sinh (Ht)
\label{solution2}
\eeq 
satisfy
\beq
\frac{\phi^2}{H^2 a^2-1}=\phi_0^2={\rm const}.
\eeq
These solutions are parametrized by $\phi_0$ and Eq.(\ref{Pphi}) gives a probability distribution for this parameter:  
\beq
dP\propto d\phi_0 \exp\left(-\frac{16\pi H^2 \phi_0^2}{3}\right).
\label{Ptilde}
\eeq
This distribution can be interpreted as describing an ensemble of $(1+1)D$ universes that nucleate with ${\tilde a}\approx H^{-1}$, $\phi\approx 0$ and $\dot\phi\approx H\phi_0$ and then evolve according to Eqs.(\ref{solution2}).
 Even though the approximations we used to derive the wave function break down at ${\tilde a}\gtrsim H^{-2}$, 
 the classical solutions become increasingly accurate at large ${\tilde a}$ and we expect the distribution (\ref{Ptilde}) to remain accurate as well.}

\section{Discussion}

Our main goal in this paper was to define and calculate the Hartle-Hawking wave function $\Psi_{HH}$ in a $(1+1)$-dimensional minisuperspace JT model with a cosmological constant $\Lambda=H^2>0$.  This model is closely related to that of a homogeneous $4D$ universe with the same cosmological constant and having spatial topology $S^1\times S^2$ (the KS model).  Our approach was first to find $\Psi_{HH}$ for the KS model using its definition in terms of a Euclidean path integral and then to use the exact correspondence between the two models to define $\Psi_{HH}$ for JT gravity.  
In our analysis of KS quantum cosmology we followed the work of Halliwell and Louko \cite{HL}.  However, this work was mostly limited to the case of a vanishing $\Lambda$, so to implement our program we had to tackle the nontrivial task of extending it to $\Lambda>0$.

 The wave function that we found is normalizable, so we could obtain a normalized probability distribution for the dilaton in the JT model.  Note, on the other hand, that the leading-order semiclassical wave function found by Maldacena {\it et al} \cite{Maldacena} is not normalizable, even after including a Schwarzian prefactor.

 Our wave function is different from the exact WDW solution obtained earlier by Iliesiu {\it et al} \cite{Iliesiu}.  This difference is due to a different choice of the boundary conditions.  The HH wave function was originally defined as a path integral over smooth Euclidean geometries with a single boundary.  We adopted this definition here and imposed the condition of smooth closure at $a=0$, where $a$ is the radius of $S^1$.
On the other hand, Ref.\cite{Iliesiu} imposed boundary conditions at large $a$  requiring that $\Psi_{HH}$ has
the asymptotic form suggested by Schwarzian theory, which accounts for quantum fluctuations of the boundary curve.  Both boundary conditions seem to be reasonable, but it appears that they are not compatible with one another. 

 The wave function obtained in \cite{Iliesiu} using the Schwarzian boundary conditions has a strong singularity at $a=H^{-1}$.\footnote{An alternative approach, suggested in Ref.\cite{Maldacena}, was to derive $\Psi_{HH}$ by analytic continuation from JT gravity with a negative cosmological constant.  A Euclidean dS metric (with $H=1$) is 
$ds^2=(1-r^2) d\theta^2+(1-r^2)^{-1} dr^2$.  With $r=\cosh\rho$, this becomes $ds^2=-d\rho^2-\sinh^2 \rho d\theta^2$, which is minus Euclidean AdS metric.  This method gives the same wave function as the Schwarzian boundary condition \cite{Iliesiu}.  We note that the origin $\rho=0$ in AdS analytically continues to the horizon ($r=1$) in dS.  This may explain why the transition amplitude from "nothing" ($\rho=0$) to some $\rho >0$ in AdS could be related to the transition amplitude from the horizon ($a=H^{-1}$) to some $a>H^{-1}$ in dS.}  
We found that this wave function is not a solution of the WDW equation.  Instead, it satisfies an equation with a singular source at $a=H^{-1}$.  Furthermore, we showed that this wave function is closely related to the transition amplitude in the KS model from $a=b=H^{-1}$ to specified values of $a$ and $b$ at the boundary, where $b$ is the radius of $S^2$.
Since a state with $a=b=H^{-1}$  (which corresponds to $a=H^{-1},~~\phi=0$ in the JT model) can hardly be interpreted as "nothing", we believe that the wave functions discussed in \cite{Iliesiu} cannot be interpreted as the HH wave function.  On the other hand, the wave function we found in the present paper satisfies the WDW equation and has only a mild singularity at $a=H^{-1}$, which one always expects in a WKB wave function at a classical turning point.  It seems therefore that our choice of boundary conditions yields a more reasonable result for the HH wave function.

A possible reason why the Schwarzian boundary condition at large $a$ fails to yield a suitable candidate for the HH wave function is that it leaves the geometry at small $a$ completely unconstrained.  
It is assumed that the geometry closes off at $a=0$ in a nonsingular way.  However, this condition is not explicitly enforced, so one should not be surprised if geometries included in the path integral include conical singularities and even gaps.


It is perhaps not surprising that the wave function we found using the boundary condition of smooth closure does not exhibit Schwarzian asymptotic behavior.  An obvious reason is that our analysis was restricted to minisuperspace, so the Schwarzian degrees of freedom were not included.  On the other hand, Iliesiu {\it et al} \cite{Iliesiu} argued that the minisuperspace wave function is simply related to the wave functional of full JT gravity. This issue needs to be better understood. Another possibility is that the condition of smoothness (absence of a conical singularity) at $a=0$ is too restrictive.  We know that the metrics contributing to the path integral are generally rather irregular, so the Hartle-Hawking proposal of integrating over smooth metrics should not be taken too literally. Finally, the Schwarzian boundary condition was imposed in Ref.\cite{Iliesiu} assuming that the dynamics of the boundary curve at large $a$ are completely decoupled from the geometry at small $a$.  It is conceivable, however, that the decoupling is incomplete, so the conditions of closure and maybe smoothness modify the asymptotic behavior of $\Psi_{HH}$. 

In this paper we utilized the familiar $4D$ minisuperspace framework in order to explore the closely related JT quantum cosmology and to define the corresponding HH wave function.  Due to the simplicity of JT theory, one can hope that with better understanding the relation between the two models will be reversed and JT cosmology will provide important insights for the $4D$ case.  Towards this goal, it would be interesting to do a path integral calculation of $\Psi_{HH}$ directly in JT model, without a reference to KS and without using the minisuperspace truncation.  This would help to understand the Schwarzian issue that we referred to above.  It would also be interesting to define the tunneling wave function in both JT and KS models.  We hope to return to these problems in future work.


\section*{Acknowledgements}

We are grateful to Jose Blanco-Pillado, Jonathan Halliwell, Oliver Janssen, Jorma Louko, Juan Maldacena and Sandip Trivedi for very useful discussions and to Raymond Laflamme for sending us his 1986 thesis.

\section*{Appendix: Higher order corrections}

The rescaled action of Eq.(\ref{SE1}) can be split into two components. The first is $\tilde{S}_{E0}=\tilde{S}_{E}(bH=1)$ and the second depends linearly on $(1-H^{2}b^{2})$. Specifically, using the definitions for $\tilde{S}_{E}, \tilde{N} , u $ and defining $v=H^{2}b^{2}$ we can decompose the action in the following way:
\beq
 \tilde{S}_{E}=\left(-\frac{2\tilde{N}}{3}-\frac{u}{\tilde{N}}-\frac{(\tilde{N}^{2}+3u)^{2}}{12\tilde{N}^{2}(\tilde{N}-3)}\right)+\left(\frac{\tilde{N}}{3}-\frac{u}{\tilde{N}}-\frac{(\tilde{N}^{2}+3u)^{2}}{12\tilde{N}^{2}(\tilde{N}-3)}\right)\left(v-1\right)   
\tag{A.1} \label{Ssplit}
\eeq
Note that this expression is exact and not an expansion to first order in $(1-H^{2}b^{2})$.

Setting $v=1$ in the action (\ref{Ssplit}) and taking a derivative with respect to $\tilde{N}$, we obtain Eq.(\ref{SSol}) for the saddles. These are the saddles of the zeroth order action $\tilde{S}_{E0}$. We will refer to the 5 solutions of this equation as $\tilde{N}_{i}$ with $i=\{1,2,3,4,5\}$.

It can be shown that if we introduce a perturbation $x$ as in Eq.(\ref{Spert}), ${\tilde N}={\tilde N}_i+x$, the action is extremized for
\beq
x=\left(\frac{1-v}{2v}\right)\frac{1}{f(\tilde{N}_{i},u)}+\frac{1}{v}h(\tilde{N}_{i},u) \tag{A.2}
\eeq
where the function $h$ vanishes at all saddles $\tilde{N}_{i}$. Thus the perturbed saddles will be given by 
\beq
\tilde{N}_{i}^{\star}=\tilde{N}_{i}+\left(\frac{1-v}{2v}\right)\frac{1}{f(\tilde{N}_{i},u)}      \tag{A.3}
\eeq
Setting $\tilde{N}=\tilde{N}_{i}^{\star}$ in the action (\ref{Ssplit}) and expanding to 2nd order in $(1-v)$ we notice the following.  The 0-th order term does not depend on the function $f$, as expected. The 1st order term takes the form
\beq
\left[\left(\frac{\tilde{N_{i}}}{3}-\frac{u}{\tilde{N}_{i}}-\frac{(\tilde{N}_{i}^{2}+3u)^{2}}{12\tilde{N}_{i}^{2}(\tilde{N}_{i}-3)}\right)+\frac{3}{8f(\tilde{N}_{i},u)}\frac{\left(\tilde{N}_{i}^{2}-2\tilde{N}_{i}+u\right)\left(\tilde{N}_{i}^{3}-4\tilde{N}_{i}^{2}-3\tilde{N}_{i}u+6u\right)}{(\tilde{N}_{i}-3)^{2}\tilde{N}_{i}^{3}}\right](v-1)    \tag{A.4}
\eeq
From (\ref{SSol}) we see that the term depending on $f$ vanishes. Thus, the action up to first order corrections is
\beq
\tilde{S}_{E}=\left(-\frac{2\tilde{N_{i}}}{3}-\frac{u}{\tilde{N}_{i}}-\frac{(\tilde{N}_{i}^{2}+3u)^{2}}{12\tilde{N}_{i}^{2}(\tilde{N}_{i}-3)}\right)+\left(\frac{\tilde{N}_{i}}{3}-\frac{u}{\tilde{N}_{i}}-\frac{(\tilde{N}_{i}^{2}+3u)^{2}}{12\tilde{N}_{i}^{2}(\tilde{N}_{i}-3)}\right)\left(v-1\right) +\mathcal{O}\left[(1-v)^{2}\right]   \tag{A.5}
\eeq
This equation is the same as Eq.(\ref{Ssplit}) with $\tilde{N}=\tilde{N}_{i}$. This means that the 1st order corrections to the action are obtained by finding the saddles for $v=1$ and inserting them into the action for $v\neq1$. The perturbed saddles contribute only to 2nd order and higher corrections to the action. Note that we did not make any specification for which saddle we are referring to. This analysis is true for all 5 saddles.

In the classically allowed region the contributing saddles for the Hartle-Hawking solution are $N_{1},N_{2}$. In this regime the function $f(N_{1,2},a)$ is given by
\beq
f(N_{1,2})=\frac{3(1-H^{2}a^{2})}{2\pm2 i \sqrt{H^{2}a^{2}-1}\pm iH^{2}a^{2} \sqrt{H^{2}a^{2}-1}}  \tag{A.6}
\eeq
\newline
Thus, the first order correction to the saddles with respect to $\left(1-Hb\right)$ is given by
\beq
H^{2}N_{1,2}=1\pm i \sqrt{H^{2}a^{2}-1}-\left(1-Hb\right)\left(\frac{2\pm2 i \sqrt{H^{2}a^{2}-1}\pm iH^{2}a^{2} \sqrt{H^{2}a^{2}-1}}{3(H^{2}a^{2}-1)}\right)  \tag{A.7}
\eeq
and the action evaluated at $N_{1,2}$ up to second order corrections is
\beq
S_{E}(N_{1,2})\approx S_{E1}(N_{1,2})-\frac{\pi(1-Hb)^{2}}{3H^{2}}\left(\frac{2\pm2i\sqrt{H^{2}a^{2}-1}\pm iH^{2}a^{2}\sqrt{H^{2}a^{2}-1}}{1-H^{2}a^{2}}\right) \tag{A.8}
\eeq
where $S_{E1}(N_{1,2})$ is the action evaluated at $N_{1,2}$ up to first order corrections in $(1-Hb)$. From the above we can specify the coefficient function $F(a)$ in Eq. (\ref{Psi+}) as
\beq 
F(a)=\frac{\pi}{3H^{2}}\left(\frac{2\pm2i\sqrt{H^{2}a^{2}-1}\pm iH^{2}a^{2}\sqrt{H^{2}a^{2}-1}}{1-H^{2}a^{2}} \right)  \tag{A.9}
\eeq
Its real part is
\beq 
\textrm{Re}F(a)=\frac{2\pi}{3H^{2}(1-H^{2}a^{2})}  \tag{A.10}
\eeq



\section*{References}

\begin{thebibliography}{}


\bibitem{Teitelboim}


C. Teitelboim, "Quantum mechanics of the gravitational field". 
\href{https://link.aps.org/doi/10.1103/PhysRevD.25.3159}{Phys. Rev. D \textbf{25}, 3159–3179 (1982)}

\bibitem{HH}
J.~B.~Hartle and S.~W.~Hawking,
``Wave Function of the Universe,''
\href{https://link.aps.org/doi/10.1103/PhysRevD.28.2960}{Phys. Rev. D \textbf{28}, 2960-2975 (1983)}




\bibitem{Vilenkin:1987kf}
A.~Vilenkin,
``Quantum Cosmology and the Initial State of the Universe,''
 \href{ https://doi.org/10.1103/physrevd.37.888}{Phys. Rev. D \textbf{37}, 888 (1988)}

\bibitem{Vilenkin:1994rn}
A.~Vilenkin,
``Approaches to quantum cosmology,''
 \href{https://doi.org/10.1103/physrevd.50.2581}{Phys. Rev. D \textbf{50}, 2581-2594 (1994)}

\bibitem{Vilenkin:1982de}
A.~Vilenkin,
``Creation of Universes from Nothing,''
 \href{https://doi.org/10.1016/0370-2693(82)90866-8}{ Phys. Lett. B \textbf{117}, 25-28 (1982)}

\bibitem{Linde:1983mx}
A.~D.~Linde,
``Quantum creation of an inflationary universe,''
\href{https://doi.org/10.1007/bf02790571}{Sov. Phys. JETP \textbf{60}, 211-213 (1984)}

\bibitem{Rubakov:1984bh}
V.~A.~Rubakov,
``Quantum Mechanics in the Tunneling Universe,''
\href{https://doi.org/10.1016/0370-2693(84)90088-1}{Phys. Lett. B \textbf{148}, 280-286 (1984)}

\bibitem{Vilenkin:1984wp}
A.~Vilenkin,
``Quantum Creation of Universes,''
\href{https://doi.org/10.1103/physrevd.30.509}{Phys. Rev. D \textbf{30}, 509-511 (1984)}

\bibitem{Zeldovich:1984vk}
Y.~B.~Zeldovich and A.~A.~Starobinsky,
``Quantum creation of a universe in a nontrivial topology,''
Sov. Astron. Lett. \textbf{10}, 135 (1984)

\bibitem{Jackiw:1984je}
R.~Jackiw,
``Lower Dimensional Gravity,''
\href{https://www.sciencedirect.com/science/article/abs/pii/0550321385904481}{Nucl. Phys. B \textbf{252}, 343-356 (1985)}

\bibitem{Teitelboim:1983ux}
C.~Teitelboim,
``Gravitation and Hamiltonian Structure in Two Space-Time Dimensions,''
\href{https://www.sciencedirect.com/science/article/abs/pii/0370269383900126}{Phys. Lett. B \textbf{126}, 41-45 (1983)}

\bibitem{Henneaux}
M.~Henneaux,
``Quantum Gravity in Two-Dimensions: Exact Solution of the Jackiw Model''
\href{https://doi.org/10.1103/physrevlett.54.959}{Phys. Rev. Lett. \textbf{54}, 959-962 (1985)}

\bibitem{Maldacena}
J.~Maldacena, G.~J.~Turiaci and Z.~Yang,
``Two dimensional Nearly de Sitter gravity,''
\href{https://doi.org/10.1007/jhep01(2021)139}{JHEP \textbf{01}, 139 (2021)}

\bibitem{Iliesiu}
L.~V.~Iliesiu, J.~Kruthoff, G.~J.~Turiaci and H.~Verlinde,
``JT gravity at finite cutoff,''
\href{https://doi.org/10.21468/scipostphys.9.2.023}{SciPost Phys. \textbf{9}, 023 (2020)}

\bibitem{Trivedi}


U.~Moitra, S.~K.~Sake and S.~P.~Trivedi,
``Jackiw-Teitelboim gravity in the second order formalism,''
\href{https://arxiv.org/abs/2101.00596}{JHEP \textbf{10}, 204 (2021)}

\bibitem{Stanford}
D.~Stanford and Z.~Yang,
``Finite-cutoff JT gravity and self-avoiding loops,''
\href{https://arxiv.org/abs/2004.08005}{arXiv:2004.08005 [hep-th]}

\bibitem{Fabbri}
A.~Fabbri, D.~J.~Navarro and J.~Navarro-Salas,
``Quantum evolution of near extremal Reissner-Nordstrom black holes,''
\href{https://doi.org/10.1016/s0550-3213(00)00661-1}{Nucl. Phys. B \textbf{595}, 381-401 (2001)}

\bibitem{Bousso}
R.~Bousso,
``Quantum global structure of de Sitter space,''
\href{https://doi.org/10.1103/physrevd.60.063503}{Phys. Rev. D \textbf{60}, 063503 (1999)}


\bibitem{HL}

J.~J.~Halliwell and J.~Louko,
``Steepest Descent Contours in the Path Integral Approach to Quantum Cosmology. 3. A General Method With Applications to Anisotropic Minisuperspace Models,''
\href{https://doi.org/10.1103/physrevd.42.3997}{Phys. Rev. D \textbf{42}, 3997-4031 (1990)}

\bibitem{StanfordWitten}

D.~Stanford and E.~Witten,
``Fermionic Localization of the Schwarzian Theory,''
\href{https://doi.org/10.1007/jhep10(2017)008}{JHEP \textbf{10}, 008 (2017)}

\bibitem{HHH}
J.~J.~Halliwell, J.~B.~Hartle and T.~Hertog,
``What is the No-Boundary Wave Function of the Universe?,''
\href{https://arxiv.org/abs/1812.01760}{Phys. Rev. D \textbf{99}, no.4, 043526 (2019)}




\bibitem{Louis-Martinez}

D.~Louis-Martinez, J.~Gegenberg and G.~Kunstatter,
``Exact Dirac quantization of all 2-D dilaton gravity theories,''
\href{ https://doi.org/10.1016/0370-2693(94)90463-4}{Phys. Lett. B \textbf{321}, 193-198 (1994)}

\bibitem{KS}

R.~Kantowski and R.~K.~Sachs,
``Some spatially homogeneous anisotropic relativistic cosmological models,''
\href{https://doi.org/10.1063/1.1704952}{J. Math. Phys. \textbf{7}, 443 (1966)}

\bibitem{Nariai}

H. Nariai, ``On a New Cosmological Solution of Einstein’s Field Equations of Gravitation". \href{https://doi.org/10.1023/a:1026602724948}{General Relativity and Gravitation, \textbf{31(6)}, 963–971 (1999)}

\bibitem{Ginsparg:1982rs}
P.~H.~Ginsparg and M.~J.~Perry,
``Semiclassical Perdurance of de Sitter Space,''
\href{https://www.sciencedirect.com/science/article/abs/pii/0550321383906363}{Nucl. Phys. B \textbf{222}, 245-268 (1983)}

\bibitem{Bousso:1996au}
R.~Bousso and S.~W.~Hawking,
``Pair creation of black holes during inflation,''
\href{https://arxiv.org/abs/gr-qc/9606052}{Phys. Rev. D \textbf{54}, 6312-6322 (1996)}

\bibitem{Laflamme}
R.~Laflamme, 
"The wave function of a $S_1\times S_2$ universe", Ph. D. Thesis, University of Cambridge (1986).
 
\bibitem{Conradi}
H.~D.~Conradi,
``Quantum cosmology of Kantowski-Sachs like models,''
\href{https://doi.org/10.1088/0264-9381/12/10/005}{Class. Quant. Grav. \textbf{12}, 2423-2440 (1995)}

\bibitem{Anninos:2012ft}
D.~Anninos, F.~Denef and D.~Harlow,
``Wave function of Vasiliev\textquoteright{}s universe: A few slices thereof,''
\href{https://arxiv.org/abs/1207.5517}{Phys. Rev. D \textbf{88}, no.8, 084049 (2013)}


\bibitem{Hertog and Conti}
G.~Conti and T.~Hertog,
``Two wave functions and dS/CFT on S$^{1}$ \texttimes{} S$^{2}$,''
\href{https://arxiv.org/abs/1412.3728}{JHEP \textbf{06}, 101 (2015)}



\bibitem{Schulman}
Schulman, L. S. (2005). {\it Techniques and Applications of Path Integration} (Dover Books on Physics) (Illustrated ed.). Dover Publications.

\bibitem{VV}

J.~H.~Van Vleck,
``The Correspondence Principle in the Statistical Interpretation of Quantum Mechanics,''
Proc. Nat. Acad. Sci. \textbf{14}, 178-188 (1928)


\bibitem{Morette}

C. Morette, ``On the Definition and Approximation of Feynman's Path Integrals,'' 
\href{https://journals.aps.org/pr/abstract/10.1103/PhysRev.81.848}{Phys. Rev. \textbf{81}, 848 (1951)}

\bibitem{Halliwell:1989dy}
J.~J.~Halliwell and J.~B.~Hartle,
``Integration Contours for the No Boundary Wave Function of the Universe,''
\href{https://journals.aps.org/prd/abstract/10.1103/PhysRevD.41.1815}{Phys. Rev. D \textbf{41}, 1815 (1990)}

\bibitem{Feldbrugge:2017kzv}
J.~Feldbrugge, J.~L.~Lehners and N.~Turok,
``Lorentzian Quantum Cosmology,''
\href{https://arxiv.org/abs/1703.02076}{Phys. Rev. D \textbf{95}, no.10, 103508 (2017)}

\bibitem{Vilenkin:1988yd}
A.~Vilenkin,
``The Interpretation of the Wave Function of the Universe,''
\href{https://journals.aps.org/prd/abstract/10.1103/PhysRevD.39.1116}{Phys. Rev. D \textbf{39}, 1116 (1989)}










\end{thebibliography}
\end{document}